# Long-range dispersal, stochasticity and the broken accelerating wave of advance

G. S. Jacobs[a,*], T. J. Sluckin[a]

[a]*Quantitative Anthropology Group, Mathematics, University of Southampton, Southampton SO17 1BJ, United Kingdom*

**Abstract**

Rare long distance dispersal events are thought to have a disproportionate impact on the spread of invasive species. Modelling using integrodifference equations suggests that, when long distance contacts are represented by a fat-tailed dispersal kernel, an accelerating wave of advance can ensue. Invasions spreading in this manner could have particularly dramatic effects. Recently, various authors have suggested that demographic stochasticity disrupts wave acceleration. Integrodifference models have been widely used in movement ecology, and as such a clearer understanding of stochastic effects is needed. Here, we present a stochastic non-linear one-dimensional lattice model in which demographic stochasticity and the dispersal regime can be systematically varied. Extensive simulations show that stochasticity has a profound effect on model behaviour, and usually breaks acceleration for fat-tailed kernels. Exceptions are seen for some power law kernels, $K(l) \propto |l|^{-\beta}$ with $\beta < 3$, for which acceleration persists despite stochasticity. Such kernels lack a second moment and are important in 'accelerating' phenomena such as Lévy flights. Furthermore, for long-range kernels the approach to the continuum limit behaviour as stochasticity is reduced is generally slow. Given that real-world populations are finite, stochastic models may give better predictive power when long-range dispersal is important. Insights from mean-field models such as integrodifference equations should be applied with caution in such circumstances.

[*]Author for correspondence -G.S.Jacobs@soton.ac.uk

1. **Introduction**

The manner in which alleles, species and diseases spread over space is of fundamental interest to population biologists. These processes have an important impact on many evolutionary and ecological systems, and are particularly relevant in the modern world, where increasing global trade [1] and highly interconnected transport systems [2] change the dynamics of disease and species dispersal. For example, international air travel has been suggested as a major driver of the spread of disease, including the 2009 H1N1 influenza A swine flu virus pandemic [3]. Anticipating species invasions, and identifying how they might progress in such conditions, is an immediate and relevant problem.

Various models have been constructed in order to theoretically explore the dynamics of spreading populations. These guide our predictions about future genetic, demographic or disease prevalence trends, and our understanding of the history implied by current patterns. A core feature of models is whether they explicitly incorporate stochasticity. Traditional approaches tend to use deterministic approximations of the underlying stochastic process. Here, there is an assumption that over many repeats of an event with a random element the stochasticity will average out, and can be ignored without invalidating results. Such models can often be analysed mathematically, but are sometimes sufficiently complex that a computational solution is necessary.

Stochastic models are usually more computationally intensive and less analytically transparent, but accept that explicitly including the randomness of events is important. It is often unclear which approach is preferable. In the specific case of species dispersal, a finite population of organisms that move and reproduce with a degree of independence implies a finite number of dispersal events. Stochasticity at small scales can have a significant impact on larger scale behaviour, and it is possible that averaging these events has a qualitative impact on model results.

One feature of population spread that is of particular practical interest is the expected rate of invasion. Deterministic equations predict that under many conditions population expansion occurs through a wave of advance travelling at constant velocity [4]. In certain cases, however, where there is a relatively high frequency of long-distance dispersal events, this wave will accelerate indefinitely [5]. The integrodifference model that retrieves this latter result has been widely applied in modelling species dispersal [6–12]. However, the approach is deterministic, and it is not clear that the underlying stochasticity of dispersal can be ignored without causing inaccuracies. The impact of randomness on the accelerating wave of advance will therefore be the principal subject of this paper. We explore this by considering a range of stochastic models and their mean-field deterministic approximations, in which many dispersal events are described as a single average process.

**Fisher-Kolmogorov and its limitations**

Classical modelling of population spread has taken the form of reaction-diffusion equations. Here, a diffusion approximation is used to model the underlying stochastic dispersal and reproduction processes, which occur concurrently and independently of one another. This is a macroscopic approximation, obtained from the stochastic description by truncating in space or time to some finite order [13]. The paradigm is the Fisher-Kolmogorov equation [14–16]:

$$\frac{\partial n}{\partial t} = \alpha n(1 - \frac{n}{K}) + D\nabla^2 n, \qquad (1)$$

where $n$ is population density at time $t$, $\alpha$ is the maximum growth rate, $K$ is the carrying capacity (in some suitable units) and $D$ is the diffusion constant. The equation is continuous in space and time and expresses the combination of logistic growth and Fickian diffusion. The diffusion constant, or diffusivity, describes the mean square distance over which a particle diffuses per unit time given a gradient of one unit, and may be expressed in dimensions $L^2 T^{-1}$. A higher diffusion constant implies that the flow of organisms from full to empty space is easier and thus more rapid.

The use of a single parameter $D$ to represent many possible dispersal regimes follows from arguments based on the central limit theorem [17]. It is justified by the relationship between Fickian diffusion and the stochastic process underlying it, Brownian motion. We can describe this process mathematically as a random walk.

A basic random walk is a stochastic system in which the position, $x$, of a particle is iteratively updated by its jump distance, drawn from a given probability distribution. This probability distribution describes the probability of dispersal over a distance $l$ in a time interval, and is known as the *dispersal kernel*, $K(l)$. If we run many random walks with a given starting position, the distribution of the particles will spread out over time. Supposing a symmetric dispersal process, the mean position remains close to zero, but the diffusivity can be captured by the deviations around this mean. For Brownian motion, and indeed more general random walks,

$$<x(t)^2> = 2Dt, \qquad (2)$$

with the constant of proportionality defining the diffusivity. The central limit theorem prescribes that the distribution function of long-time positions is Gaussian so long as the same kernel applies to all particles, there are no long-range correlations in jump-distance, and the kernel has a finite first and second moment. $D$ is related to the variance of the kernel by

$$D = \frac{1}{2} \int_{-\infty}^{+\infty} l^2 K(l) \mathrm{d}l. \quad (3)$$

When the variance is unbounded, $D$ is similarly not well defined, a point we return to shortly.

An initially isolated population that behaves according to the Fisher-Kolmogorov equation spreads out over time, creating a 'wave of advance', while maintaining a logistically determined level of occupation behind the travelling front. The model has been subject to much mathematical investigation, and a range of velocities can be sustained. However, under suitable initial conditions [15], including those most relevant to biological invasions, the wave speed (after transient acceleration) asymptotically approaches

$$c = 2\sqrt{\alpha D}. \quad (4)$$

For $c$ to be asymptotically constant both $D$ and $\alpha$ must exist and be asymptotically constant.

Laying aside model-specific issues such as environmental heterogeneity, advection, and qualities of population growth such as Allee effects, there are two general concerns about the application of the Fisher-Kolmogorov equation. Firstly, long distance dispersal may complicate the diffusion term. Secondly, stochasticity may invalidate results obtained from averaged processes. We deal with these points in turn.

**Long-distance dispersal through integrodifference models**

Standard theory suggests the diffusivity $D$ can capture a wide range of stochastic dispersal processes through the relationship in Eq. (2). In the context of population spread, a naïve assumption of a normally distributed dispersal kernel would seem reasonable. However, many species appear not to follow this dispersal pattern, with dispersal better represented by a 'fat-tailed' kernel. These kernels involve an excess probability of dispersal at longer distances; specifically, the tail of the dispersal kernel decays more slowly than an exponential distribution. Such dispersal regimes have been observed in fungal spores [18], plant seeds [19], and in mammals and birds [20]. Under these conditions, it becomes less clear that $D$ will capture the dispersal process faithfully, and there is a strong argument for explicitly incorporating the dispersal kernel itself into a model.

As we have noted, some fat-tailed kernels decay so slowly that the variance or other moments are not well defined. Specifically, when the tail of a kernel decays as a power law, $K(l) \propto l^{-\beta}$ as $l \to \infty$, the $(\beta - 1)^{\text{nth}}$ and greater moments are not finite. This phenomenon is due to the dominant role that rare large values have on the characteristics of the distribution, and is useful for incorporating a relatively high probability of extremely long-range events into the dispersal regime. If $\beta \leq 3$, we can predict dispersal behaviour by considering a particular class of random walks, known as Lévy flights, for which the second moment is undefined [21, 22].

When the variance is unbounded the effective diffusivity increases with time, termed superdiffusion. Given the role of $D$ in the Fisher-Kolmogorov equation, we might expect these kernels to lead to an accelerating wave of advance. Indeed, reaction-diffusion equations of the Fisher-Kolmogorov type that model Lévy flights as fractional diffusion have been investigated, and lead to exponentially accelerating waves [23]. There is a significant body of theoretical and empirical work investigating Lévy flights in the context of foraging behaviour [24].

Several authors [5, 25] have pointed out that systems subject to fat-tailed kernels which nevertheless still possess a finite variance (and hence a well-defined $D$) may also exhibit anomalous behaviour at long times. This is reflected in the lack of analytic behaviour of $\tilde{K}(k)$ at low $k$, where $\tilde{K}(k)$ is the Fourier transform of the dispersal kernel $K(l)$. The low $k$ non-analyticity then leads to problems in the application of the central limit theorem at long times. To determine more accurately the implications of such anomalous kernels for species diffusion, we need to describe the dispersal process explicitly, rather than summarising it merely in terms of a diffusivity $D$.

Kot *et al* [5] achieve this mathematically by incorporating the dispersal kernel directly in an integrodifference equation of the form:

$$n(x, t+1) = \int_{-\infty}^{+\infty} K(x-y) f[n(y,t)] \mathrm{d}y. \tag{5}$$

The function $f[n(y,t)]$ applies the population growth process, while the (normalised) dispersion kernel $K(x - y)$ represents the relative probability of dispersal between positions $x$ and $y$ in continuous space, with $l = y - x$. Importantly, there is no assumption that the underlying stochastic dispersal process is Brownian. However, unlike the Fisher-Kolmogorov equation, time-steps are discrete, which can lead to velocity deviations from the continuous time case in similar systems [26, 27]. Furthermore, growth and dispersal are no longer concurrent and independent. Rather, growth and dispersal occur sequentially, such that there is a coupling between the two processes.

We can describe Eq. (5) as a mesoscopic representation of the stochastic disper-sal and growth processes, in that the random behaviour of individual organisms is averaged as a probability density function [13]. The time-evolution of the probability distribution for occupation over space is then studied, which can be considered population density when a population is large. The approach thus captures elements of organism movement that are summarised by $D$ in the macroscopic reaction-diffusion model of Eq. (1). Under certain circumstances, such as a normally distributed dispersal kernel combined with logistically limited growth, Eqs. (5) and (1) retrieve identical wave velocities [5]. However, this does not imply that the microscopic processes for which they provide deterministic approximations are identical.

In apparent contradiction to Fisher-Kolmogorov predictions, certain fat-tailed dispersal kernels with finite second moments lead to indefinitely accelerating waves of advance [5] in the integrodifference modelling framework. Given that more information is preserved about the dispersal kernel in the integrodifference approach, we might regard it as more broadly applicable. Specific kernels with this effect include stretched exponentials and power laws, where for large dispersal distance $l$, $K(l) \propto e^{-|l|^\gamma} : \gamma < 1$ and $K(l) \propto |l|^{-\beta} : \beta > 3$ respectively. The stretched exponential kernel leads the spatial extent of the wave to increase as a power law over time, with exponent $\frac{1}{\gamma}$ [13]. Evidence from both reaction-diffusion equations [23, 28] and integrodifference models [5, 29] indicate that power law kernels cause wave velocity to increase exponentially with time, an effect that persists when $\beta > 3$ [28].

There are many documented examples of apparently accelerating species invasions (eg. rice water weevil, *Lissorhoptirus oryzophilus*, in Japan, [30]; cheatgrass, *Bromus tectorum L.*, in North America, [31]; among other diverse plant species, see [32]; potentially Californian sea otters, *Enhydra lutris nereis*, [11]; also see [33]). In some cases, these behaviours may be due to factors other than long-range dispersal. Nevertheless, the link between long distance events and accelerating waves has been explicitly suggested with respect to the spread of several plant and human pathogens using data from empirical studies and and observed invasion events [34]. In these cases dispersal is either by wind or via avian vectors.

As Kot *et al* noted, indefinite acceleration is biologically unsustainable, and can break down for several reasons. These include the introduction of an Allee effect, a long-distance cut-off to the dispersal kernel, and effectively introducing a spatially determined cut-off to dispersal by limiting system size [5]. Stochasticity can also have a pronounced effect on system behaviour.

Although the implications of certain of these features remain unclear, integrodifference equations are frequently used in species invasion modelling. The flexibility afforded by explicitly representing the dispersal kernel has allowed authors to explore various phenomena, often incorporating long-distance dispersal [5–12, 29, 35].

Recently, the approach has been suggested as one of four preferred methods for pest risk analysis [36]. It must be emphasised that models in general, and here the integrodifference equations method in particular, are only an approximation of real-world behaviour, and if the implicit assumptions are incorrect the results will also be unreliable.

**Demographic stochasticity: ambiguous results**

Both the descriptions of population spread introduced above are deterministic. They are justified by the belief that they will capture the essential behaviour of the underlying stochastic processes of reproduction and dispersal. Under which circumstances they are the correct deterministic limits is unclear.

Demographic stochasticity is known to have an impact on model behaviour. A reduction in wave velocity is usually suggested [37–39], though contradictory results exist for a two dimensional stochastic cellular automata model [40]. In the simple, linear case, where population growth and dispersal are not density-dependent, work with branching random walks suggests that introducing demographic stochasticity does not generally slow invasions [41]. Separately, the interaction between dispersal kernel and reproductive rate has been highlighted as having an important impact on the structure of the wave [42].

In the context of the non-linear Fisher-Kolmogorov equation, a productive route of enquiry has been approximating stochastic effects by introducing a cut-off to the population growth term, thereby reducing wave velocity [43]. This approach has allowed the characterisation, for example, of stochasticity-induced velocity corrections where the dispersal kernel is exponential, representing the boundary case past which the Fisher-Kolmogorov approach cannot be naïvely applied [44]. In cases where acceleration is predicted by deterministic models, demographic stochasticity has been suggested to break acceleration, even given extremely fat-tailed kernels with unbounded variance [37, 39]. However, noting Lévy flight predictions and Mollison's density-dependent model of epidemic spread in continuous time [25], which preserves acceleration in the context of these kernels, results remain ambiguous.

In summary, models that are able to incorporate long-distance dispersal, demographic stochasticity due to finite population size, and density-dependence (which has a deep theoretical history, [45]) are most biologically plausible but least understood. It is unclear, for example, when exactly wave acceleration should be broken by stochasticity in dispersal or reproduction, and how this might occur. To explore this in more detail, we here present the results of simulation modelling of population spread incorporating two well-known classes of dispersal kernel. Although explicitly incorporating stocahsticity into models is intuitively most relevant when predicting the spread of species with lower population size/fewer dispersal events, we find that doing so may be prudent in many situations, and particularly when dispersal is best represented by a fat-tailed kernel.

The layout of the paper is as follows. In §2 we introduce the modelling framework of our simulations. In §3 we present results for both stochastic and deterministic models, finding that most waves which accelerate in the mean-field model do not when stochasticity is introduced §3.3, 3.2. However, acceleration persists in the stochastic model for power law kernels when $\beta < 3$. We also examine the impact of reducing stochasticity by increasing the carrying capacity. Here, the behaviour of the finite-population, stochastic model only approaches that of the mean-field system slowly when dispersal is represented by fat-tailed kernels §3.4. This has important implications regarding the situations in which averaging approaches such as integrodifference equations may be successfully applied. In our final simulations, we consider the effect of a long-distance kernel cut-off on wave velocity and dispersal dynamics §3.5. This represents an interesting case as truncated dispersal kernels, particularly power laws, are often encountered in the literature [46–48]. We conclude by confirming our results for several variations on our model, §3.6, and discussing our results in the context of previous work, §4.

## 2. Simulation modelling

We here introduce our two dispersal models, a stochastic model and its mean-field approximation, and describe our methods of data analysis and the simulations conducted. The stochastic model design builds on the *simple epidemic* model explored by Mollison [25] and also bears some resemblance to the dispersal model of Clark *et al* [37] and to Kot *et al's* linear branching random walk model for population spread [41]. Our models incorporate density dependence in the dispersal process, and follow an algorithm that might be described as a 'seeding random walk'. As with the models of Mollison and of Clark *et al*, dispersal and reproduction are united as a single process. As a result, they more closely resemble the emission of seeds by plants than animal dispersal, in which movement is coupled with reproduction. Given this, we also briefly discuss the differences between our model of dispersal and the Fisher-Kolmogorov population diffusion model.

All invasions occur across a 1-dimensional lattice of size $L$ and are discrete in both space and time. Initial conditions have all sites unoccupied apart from the left-most site, $n(x = 0, t = 0) = 1.0$ and $n(x > 0, t = 0) = 0.0$. A dispersal/growth process is iterated through time, and simulations are terminated when system filling, $n(L)$, reaches 90% ($n(L) = \sum_{x=0}^{L} \frac{n(x,t)}{L} \geq 0.9$). This termination condition implicitly defines our measure of wave velocity, which is estimated using the time until termination for different size systems. For simulations incorporating death we apply other measures of population extent, see §3.6.

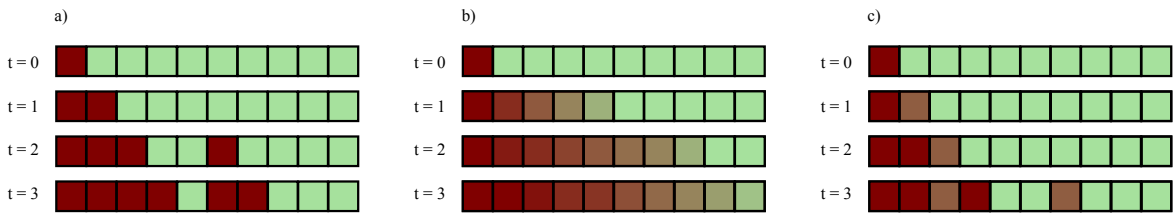

Figure 1: Diagram of simulation design and example evolution for the models explored. a) The stochastic Model 1, with $N = 1$; b) the mean-field deterministic Model 2 and c) the stochastic Model 1, with $N = 2$ and a resultant reduction in stochasticity.

We ignore dispersal events to sites outside of the lattice. Such boundary conditions are biologically reasonable in many cases. However, they lead to two possible concerns. Firstly, there is the possibility of extinction by over-dispersal in some models [16, 49]. We generally ignore death in our simulations in order to focus on stochasticity associated with dispersal, so this is rarely a problem. Secondly, a finite-size system limits the maximum distance that a propagule can travel. As we repeat simulations with increasing system size, an artefactual increase in the distance to which the dispersal kernel stretches occurs. This in turn implies a change in the rate of reproduction. Given these concerns, we explore periodic boundary conditions (see Appendix 2), with no qualitative change to results.

**Stochastic Model 1**: We begin by describing our stochastic model algorithm, represented in Fig. 1a and 1c. Occupation at each lattice site, $n(x,t)$, is a value $\frac{n}{N}$, with integer $n : 0 \leq n \leq N$. After setting up the initial conditions, we proceed through a series of iterated steps:

1. The population of each site *reproduces*, making $n(x,t)N$ reproduction attempts with successful birth probability $b$ for each event.
2. Each newborn individual *disperses* a random distance $l$ drawn independently from the dispersal kernel $K(|l|)$, with a logistic probability of success $1 - n(x + l, t)$.
3. Once each reproduction/dispersal event has occurred, the population of each site is *updated* according to the number of successful propagules arriving at it. Each successful propagule increases site filling by $1/N$.

4. *Death* is now implemented, each individual dying with a probability $d$, such that there are up to $n$ deaths in a site with filling $n(x,t) = n/N$.
5. For any site with filling $n(x,t) > 1.0$, the population is reduced to $n(x,t) = 1.0$.
6. Steps 1 to 5 are repeated until the termination condition is met, or population extinction occurs.

Dispersal can occur in either direction. When $N = 1$, dispersal is highly stochastic, each occupied site making a single dispersal attempt per generation. As $N$ is increased, a fully occupied site releases increasingly many propagules of decreasing size. Stochasticity decreases, and we approach the mean-field model presented below; we conjecture that in the limit $N \to \infty$ we indeed recover this mean-field model. In these models, stochasticity arises both in the dispersal/birth process and through death. However, here we focus on the randomness of dispersal, and usually set $b = 1$ and $d = 0$. Our main results are found to be robust to positive $d$ and $b < 1.0$, as explored in §3.6.

**Mean field Model 2**: This is an average deterministic representation of Model 1, represented in Fig. 1b. We now follow a mean-field approach, in that we assume that the dispersal of very many interacting propagules can be described in terms of a single average process. The occupation of each site can now be defined so that $0.0 \leq n(x) \leq 1.0$. After setting up initial conditions, we update the system iteratively using our mean-field equation,

$$n(x, t+1) = (1-d)n(x,t) + (1-d)b[1 - n(x,t)] \sum_{l=-\infty}^{+\infty} K(|l|)n(x+l, t), \quad (6)$$

where $d$ and $b$ are considered the average effect of the birth and death probabilities in Model 1, with $0 \leq d, b \leq 1.0$. In this equation, site occupation at $t+1$ is determined by a logistically limited dispersal from all other sites, with birth rate $b$, followed by a death stage. The model resembles the integrodifference version of the simple epidemic explored by [35], though differences exist in our application of death and in the lattice structure. Given that birth and dispersal are not always commutative in reaction-diffusion systems [50], these differences may be important. As such, we intend this equation as a tool to investigate our stochastic algorithm rather than a general representation of most dispersal processes.

Note that this is not a spatially discretised version of the integrodifference equation Eq. 6 in Kot *et al* [5]. In that model, a potentially non-linear population growth function is followed by dispersal that is not logistically limited. Integrodifference models of this form are popular in the literature and more closely reflect the life history certain organisms than our models. We therefore confirm our main results in a lattice variant of this integrodifference system in §3.6.

**Dispersal kernels**: To represent complex dispersal regimes, we follow Kot *et al* [5] in describing a dispersal kernel. Our models are discrete in space, and the probability of dispersal between sites $x$ and $y$, over a distance $l = x - y$, is:

$$K(|l|) \propto f(|l|), \qquad l \in \mathbb{Z}, l \neq 0 \tag{7}$$

We define $K(0) = 0$ in all cases, in order to avoid an infinite probability of zero-length jump distance for power law kernels.

To represent kernels incorporating long-distance dispersal, we use two classes of function that can lead to fat-tailed distributions:

$$f_a(l) = e^{-|l|^\gamma}, \tag{8}$$

$$f_b(l) = |l|^{-\beta}, \tag{9}$$

with $l \neq 0$ and $l \in \mathbb{Z}$ in both cases. When functions $\{f_a\}$ or $\{f_b\}$ apply for $|l| \to \infty$, they respectively describe the stretched exponential (if $\gamma < 1.0$) and inverse power law functions. In the real world, short-range behaviour may deviate substantially from these idealised distributions. However, such deviations are unlikely to impact long-time invasion behaviour. For this reason, we also expect that distortions to our dispersal kernels in the model, due to the discrete lattice or constraining $K(0) = 0$, to have minimal impact on qualitative system behaviour.

Our chosen kernel forms offer flexibility in investigating wave of advance behaviour. For functions $\{f_a\}$, $\gamma = 1$ is an exponential distribution and $\gamma = 2$ is a Gaussian. Functions $\{f_b\}$ with $\beta \leq 1 + z$ lack finite moments greater or equal to the $z^{\text{th}}$ moment. In one dimension, kernels with $\beta \leq 3$ do not have finite variance, leading to interesting behaviour. These are the kernels leading to Lévy flights.

Having determined the relative probability of a dispersal at each distance, we normalise to obtain the kernel:

$$\sum_{-l_{\max}}^{+l_{\max}} K(|l|) = 1. \tag{10}$$

Note that taking the absolute value of $l$ in the above formula corresponds to symmetrical dispersal behaviour in both directions. Although the kernel should theoretically extend to $l_{\max} = \infty$, for practical reasons we define a cut-off distance over which dispersal cannot occur. This is set to $l_{\max} = 2 \times 10^8$, far larger than most of the invasion lattices explored, and can be regarded as a sum out to infinity.

Some caution must be applied here. One needs to have $l_{\max}$ large, but small enough for computer memory considerations. In the case of power-law kernels (9) we can check the validity of this approach through an alternative method of constructing the kernel, using the Hurwitz zeta function [51]. This check sometimes indicated an artefactual increase in filling time for very large systems ($L \geq 10^7$), particularly using long range power law kernels, $\beta < 3$. In some of our simulations we investigate the impact of a cut-off to dispersal distance, by reducing $l_{\max}$ to values less than the system size and normalising as above, §3.5.

**Comparison of Models**: The above models share various essential features. Specifically, they are discrete in space and time and incorporate dispersal and growth through a logistically limited "budding" process. Dispersal distance is determined according to a dispersal kernel. Initial conditions, the treatment of boundaries, and termination conditions are the same. Finally, in the main body of the work all models ignore population death, other than that implicitly included when population growth is seen as a net process of death and birth. We assess the effects of including death in §3.6. In many important points the two models above are comparable.

The algorithm for dispersal and birth in our models well-represents the spread of many plants, and combining the two processes follows the approach of other workers (eg. [37] in a population dispersal context, and [25] for modelling epidemic spread). However, the differences between this and the Fisher-Kolmogorov model could lead to difficulties in interpreting our results and situating them historically. In addition to the non-Markovian characteristics of our model, whereby dispersal and reproduction no longer occur independently and concurrently, a particular difference between the Fisher-Kolmogorov model appears in the application of the logistic effect. To investigate this latter point, we give a basic derivation of a diffusive limit of our mean-field equation (6) in Appendix 1.

This derivation highlights similarities with a diffusion approximation of Mollison's simple epidemic [52], and we find that several analytic results from his model hold when we reduce the spatial and temporal scale of our system, §3.1. Our application of the logistic effect based on filling at the target site translates to a non-linearity in the diffusion term, see Eq. (A1.8), having greatest impact when the system nears filling. We nevertheless expect our mean-field results to be qualitatively quite general. The Linear Conjecture states that the speed of a wave of advance is governed by the linear properties of the governing partial differential equation far ahead of the front itself, and is thought to apply when the average reproductive rate of an individual is maximised in virgin territory and individuals have negligible impact on the environment far from their current location [53]. These conditions hold for our model. Note that the latter point refers not to long-distance dispersal, but to effects such as local population growth causing global environmental degradation [54].

In support of this argument, we find agreement between our simulations and standard wave number selection methods [55] for determining wave speed, see Fig. 2 and Appendix 3. We also find that explicit mean-field simulations in which the logistic effect is contributed by the home site yield very similar results to Model 2, although there are interesting deviations in the stochastic case under some long-range dispersal regimes, see §3.6. As a more general point, the partitioning of the logistic effect between the home range, the target of dispersal, and sites along the route of dispersal technically depends on the life-history one seeks to describe. As such details can impact deterministic and stochastic systems in different ways, there is an argument for considering them when designing models for ecological applications, especially when long-range dispersal is important.

**Methodological comments**: To investigate the impact of stochasticity on the dispersal process, we perform simulations of the two models for various parameterisations of the two types of dispersal kernel. This allows us to explore a highly stochastic scenario with $N = 1$, a deterministic scenario, and the effects of reducing randomness by increasing $N$.

To quantify the velocity behaviour of the different dispersal regimes, we use finite size scaling [56]. Here, the size of the lattice is varied over several orders of magnitude, $10^2 \leq L \leq 10^7$ where possible, and the filling time recorded. This method effectively switches the dependent and independent variables, such that we no longer have to follow the wavefront explicitly. The approach allows us to obtain reliable results given noisy systems, in which the wavefront can consist of a large, sometimes widening, region of sparse filling that is difficult to define. The behaviour of filling time as system size is increased is used to identify the relationship between the dispersion kernel and wave velocity and acceleration. Appropriate functional forms are used to interpret results in the different cases.

**Finite size scaling:** For systems in which invasion takes place through a constant velocity wave of advance, the filling time can be approximated as

$$T \approx \frac{L}{c}, c \approx \frac{L}{T}, \qquad (11)$$

where $L$ is system size, $T$ is filling time (or average filling time in stochastic systems) and $c$ is the velocity of the wave of advance.

Some accelerating waves can be approximately described by the relation

$$L \approx aT^B, \tag{12}$$

with parameter $a$ dominating early time velocity and parameter $B > 1$ describing acceleration. Log-log plots of $\ln T$ against $\ln L$ enable us to approximate the value of these parameters using a linear fitting

$$\ln T \approx \frac{1}{B} \ln L - \frac{1}{B} \ln a. \tag{13}$$

In certain systems, rapid exponential acceleration has been observed [5, 23, 28]. In such cases, filling time and system size scale as

$$T \approx g \ln L + h, \tag{14}$$

where the parameter $h$ describes early time behaviour and parameter $g$ the acceleration effect. A linear fitting to a semi-log plot of $T$ against $\ln L$ is appropriate for estimating the parameters. This model can be seen as a correction to the non-local dispersal case, where propagules disperse to random sites on the lattice and total system filling, $n(L)$, follows

$$n(L) \approx e^{\alpha t} \quad \Rightarrow \quad t \approx \frac{1}{\alpha} \ln n(L), \tag{15}$$

for early times.

These relations represent tools with which we can characterise our idealised systems in the parameter space explored, and do not necessarily correspond to analytically retrievable behaviour.

The relationships of the various variables above to the kernel parameters were estimated using the non-linear regression analysis package GraphPad Prism version 6.0 (GraphPad Software, La Jolla California USA, www.graphpad.com).

**Supplementary investigations**: We also conducted several additional investigations in order to clarify the behaviour of our simulations and facilitate comparison with the work of other authors:

a) We have verified the behaviour of our simulation model in several ways. Allowing the model to approach the limits of continuous space and time, we retrieve analytic results for the simple epidemic (§3.1). Our dispersal kernels yield expected diffusion coefficients, Fig. A3, and wave velocity given a fat-tailed kernel in our stochastic system remains within expected bounds, Fig. 5, as defined by an approach suggested in Clark *et al* [37].

b) To assess the implications of the various dispersion kernels, we directly estimated $D$ using a random walk simulation and Eq. (2). The relationship between $D$ and wave velocity is explored in Appendix 4.

c) We also investigated the impact of stochasticity in birth and death, a stochastic model following the integrodifference equation of Kot *et al*, and a version of Model 1 with the logistic effect acting on propagules but arising at the home site, §3.6. These models confirm our main results.

d) To clarify the behaviour of potentially more realistic dispersal scenarios, simulations were conducted using the bivariate Student's $t$ dispersion kernel, as recommended and investigated by Clark *et al* [37], Appendix 6. Our main results also hold qualitatively when using different methods of measuring wave velocity, eg. Fig. 10, and when applying periodic boundary conditions, Appendix 2.

## 3. Results

One of our key observations will be that, when long-range dispersal is important, demographic stochasticity gives rise to qualitative differences in wave velocity behaviour, as compared to predictions from integrodifference modelling [5] and our own mean-field model (See Fig. 3 and Table 1). We find that stochasticity breaks wave acceleration caused by fat-tailed kernels, except in the case of power law kernels with $\beta < 3.0$. This critical point supports some previous results [25] but deviates from expectations based on the behaviour of similar systems in which stochasticity also disrupts acceleration in the Lévy flight case [37, 39].

The difference in wave velocity behaviour between stochastic and deterministic models, which can persist even when $N$ is large (see Figs. 6, 8), suggests that a realistic representation of demographic stochasticity is important in models of species invasions incorporating long-range dispersal.

*3.1. Testing model behaviour*

We applied several tests to confirm that our programs display expected model behaviour. Firstly, we obtained the asymptotic velocity of our mean-field Model 2 for several simple kernels (nearest-neighbour: $K(l = \pm 1) = 0.5$, $K(l \neq \pm 1) = 0.0$; normal distribution; exponential distribution). When $b = 1$ and our standard spatial scale is applied, these deviate somewhat from both Fisher-Kolmogorov expectations and those for the infinite-population limit of the simple epidemic in continuous space and time (see [52], Table 1).

We therefore explored the impact of increasing the resolution of the spatial lattice, by defining our kernels as $K(|l_\varphi|)$, with $l_\varphi = \dfrac{l}{\varphi}$, $\varphi l \in \mathbb{Z}$, and of the time steps, by reducing $b$. Here, $b$ serves as a temporal scale rather than a birth rate, so we

normalise the resulting velocity according to $\frac{1}{b}$. Results obtained for the simple epidemic model with a nearest neighbour dispersal kernel are approached as $b \to 0.0$, see Fig. 2a. For the normal and exponential distributions, analytic results are approximated when $b \ll 1.0$ and $\varphi \gg 1$. These findings are unsurprising, as low-$b$, large $\varphi$ systems more closely resemble models that are continuous in time and space, and the simple epidemic model structurally resembles our system.

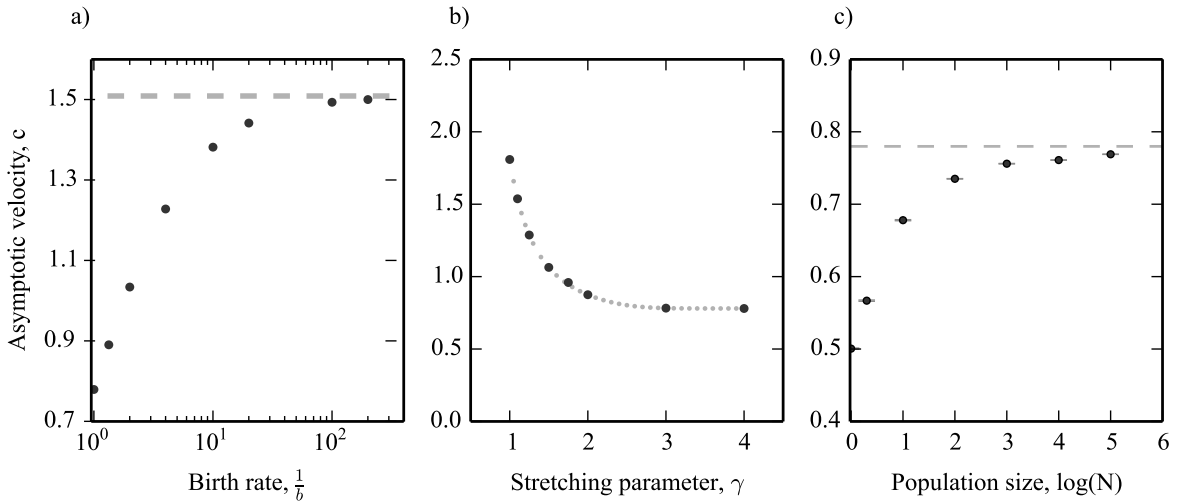

Figure 2: Testing model behaviour: a) Approach to expected asymptotic velocity for the nearest-neighbour kernel ($D = 0.5$) as birth rate, $b$, is reduced, with velocity rescaled as described in the main text; b) Agreement between asymptotic velocity obtained by simulation and expected velocity obtained using the marginal stability of the linearised wave front for exponential family kernels, $\gamma \geq 1.0$; c) Approach to mean-field (Model 2) asymptotic velocity as $N$ is increased in Model 1, thereby reducing demographic stochasticity, using the nearest neighbour kernel and 5-20 replicates. Relative standard errors are plotted, but are minimal.

We also applied a well-known argument based on the marginal stability of the linearised form [4, 55, 57] of the population function far ahead of the wave front, see Appendix 3. This allowed us to correctly estimate the wave velocity for the nearest-neighbour ($c \approx 0.78$) and exponential family of kernels ($\gamma \geq 1.0$) in the mean-field Model 2 case, both for $b = 1.0, \varphi = 1$ (see Fig. 2b) and when $b < 1.0, \varphi > 1$.

Finally, we confirmed the convergence of wave velocity to the mean-field Model 2 result as stochasticity is reduced by increasing $N$. We use the nearest-neighbour kernel, Fig. 2c, on an lattice of size $L = 10^4$.

The difference between asymptotic wave velocity and the Fisher-Kolmogorov prediction suggests that the simple relationship between diffusion constant and velocity, $c = 2\sqrt{\alpha D}$, may not hold. As results are not critical to this study, we offer a basic investigation of this in Appendix 4. Briefly, a relationship of the form $c = \mu D^\rho$ is apparent, with $\rho$ moderately close to 0.5, for all cases where the kernel leads to a constant velocity wave. However, some deviations exist. These appear to be structural, such that reducing $b$ and increasing $\varphi$ does not negate them, in agreement with results in Mollison [52] on the simple epidemic. For example, simulations with $b = 0.01$ and $\varphi = 1$ yield $c \approx 2.26 D^{0.59}$ for the mean-field model with the exponential family of kernels, $\gamma > 1.0$. We can also quite accurately approximate the impact of discrete time and space using the linearisation method. This allows us to estimate the relationship between $c$ and $D$ for the $b = 0.00001, \varphi = 10^5$ system, which would require lengthy simulations, as $c \approx 2.599 D^{0.578} \approx \dfrac{3\sqrt{3}}{2} D^{\frac{1}{\sqrt{3}}}$.

*3.2. Model 1 with $N = 1$: Constant velocity waves in a stochastic lattice system, with a notable exception*

In the highly stochastic Model 1, an asymptotically constant wave velocity is generally observed, apparent in the linear scaling of filling time with system size (Figure 3a, c). This is usually true even when long-distance dispersal is incorporated through a fat-tailed dispersal kernel, contrasting with predictions from integrodifference equations [5] and our own deterministic model (see below). An accelerating wave of advance is seen, however, for kernels that lack a finite second moment. Approximate velocity behaviours obtained by non-linear regression on filling time behaviour or velocity are given in Table 1.

For the exponential family of kernels, $K(l) \propto e^{-|l|^\gamma}$, a constant velocity wave is observed even when the kernel is fat-tailed ($\gamma < 1.0$). So long as $\gamma > 0$, the diffusivity $D$ remains well-defined. Velocity as determined through our simulations (Fig. 4) was exceptionally large when $\gamma$ is small. Our results can be quite closely approximated using the theoretical diffusion constant [58] of the stretched exponential kernel (Appendix 4), $D \approx \dfrac{\Gamma\left(\frac{3}{\gamma}\right)}{\Gamma\left(\frac{1}{\gamma}\right)}$ where $\Gamma$ is a gamma function.

Simulations using a power law kernel, $K(l) \propto |l|^{-\beta}$, show a constant velocity when $\beta > 3.0$, Fig. 3c. However, when $\beta < 3.0$, accelerated invasion fronts persist, as predicted by [25]. Acceleration is not exponential as in our mean-field Model 2 (see §3.3). Rather, the linear fit in Fig. 3c suggests acceleration occurs as a power law, as in the case of the stretched exponential kernels in Model 2.

Estimates of Lévy flight behaviour [39, 59] yield

$$\left(<x(t)^2>\right)^{1/2} \approx t^{1/(2-\beta_2)}, \tag{16}$$

where $\beta_2 = 3.0 - \beta$. The diffusion constant for such distributions is undefined. However, in a single generation we take a finite number of samples from the kernel, and the variance is finite. We might expect the diffusion constant to increase with system filling, then, implying the acceleration effect that we observe. We were able to estimate an expression for velocity behaviour when $\beta$ is not too small, $2.95 \geq \beta \geq 2.15$, which supports a strong dependence of velocity on time, see Table 1.

| Kernel form | Model 1 (stochastic, $N=1$) | | Model 2 (mean-field) | |
|---|---|---|---|---|
| | Equation | Behaviour | Equation | Behaviour |
| Exponential, $\gamma > 1.0$ | $0.5 + (0.52 \pm 0.01)\gamma^{-3.49 \pm 0.12}$ | Asymptotically constant | $0.78 + (1.03 \pm 0.03)\gamma^{-3.23 \pm 0.19}$ | Asymptotically constant |
| Exponential, $\gamma < 1.0$ | Inaccurate fitting | Asymptotically constant | $\frac{(0.94 \pm 0.03)a_1}{\gamma^{1.10 \pm 0.05}} t^{\left(\frac{0.94 \pm 0.03}{\gamma^{1.10 \pm 0.05}} - 1\right)} \approx \frac{a_1}{\gamma} t^{(\gamma^{-1} - 1)}$ | Accelerating |
| Power law, $\beta > 3.0$ | $0.5 + (0.14 \pm 0.01)\beta_1^{-1.09 \pm 0.06}$ | Asymptotically constant | $g^{-1} \exp(\frac{t}{g})$ | Rapidly accelerating |
| Power law, $\beta < 3.0$ | $a_2(1.0 + B)t^B$, where $B \approx (3.21 \pm 0.26)\beta_2^{1.59 \pm 0.23}$ | Accelerating | $g \approx (1.45 \pm 0.02)\beta - (0.04 \pm 0.16)$ | Rapidly accelerating |

Table 1: Table showing wave velocity behaviour given exponential, $K(l) \propto e^{-|l|^\gamma}$, and power law, $K(l) \propto |l|^{-\beta}$, dispersal kernels, as estimated using non-linear regression on finite-size scaling simulation results. These correspond to observed behaviour in our simulations and do not necessarily describe analytically retrievable relationships. $\beta_1 = \beta - 3$ and $\beta_2 = 3 - \beta$, while $a_1$ and $a_2$ are prefactors for the velocity behaviour - these and tend to be important at early times. We find that $a_1 \approx \frac{0.22 \pm 0.07}{(1.0 - \gamma)^{0.96 \pm 0.16}}$. We could not obtain an accurate fitting for $a_2$, which was $< 10^{-3}$ when $\beta \leq 2.25$ and very small indeed when $\beta < 2.0$, suggesting an important influence even when $t$ is relatively large. We discuss this briefly in the main text. We fit this regime in the region $2.95 \leq \beta \leq 2.15$. For more details on the parameter ranges for which each fitting was retrieved, see Appendix 4 and Fig. A4.

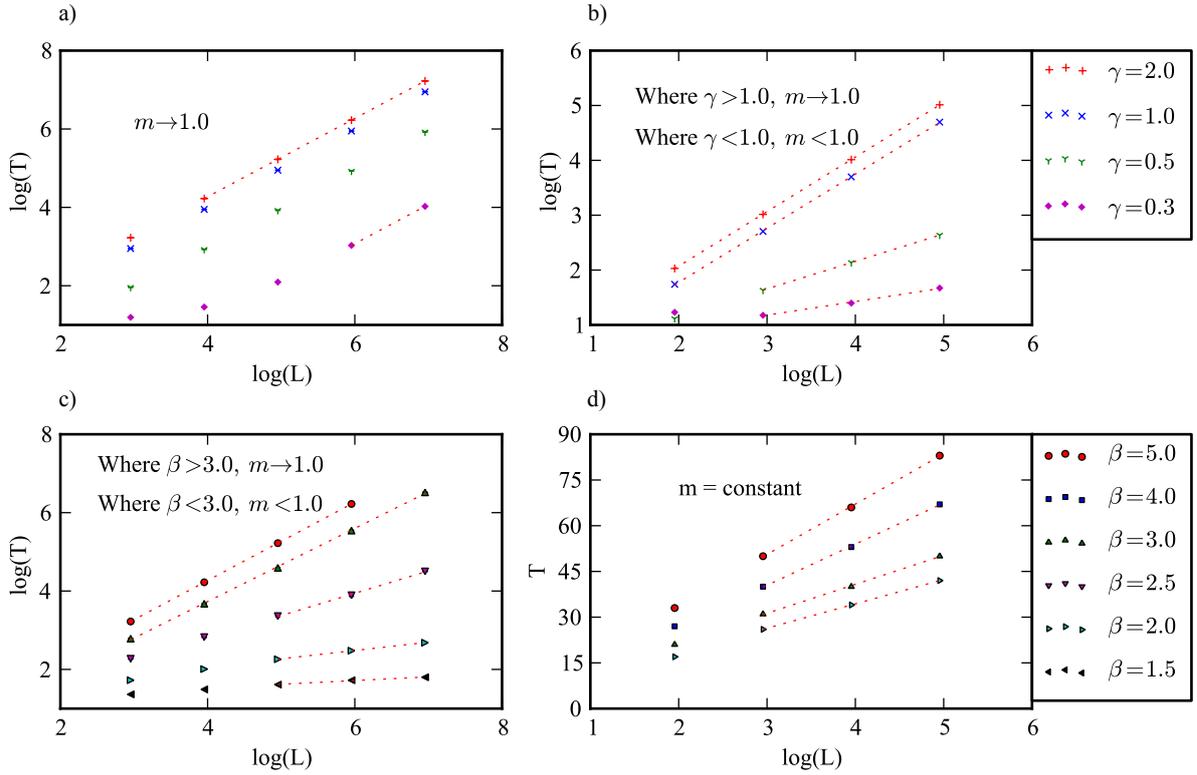

Figure 3: Log-log/semi-log plots of filling time, $T$, against system size, $L$, for the two kernel families following stochastic and mean-field systems. $m$ indicates the gradient of a linear fitting to the data, with $m < 1.0$ indicative of acceleration in the log-log plots. Top-left (a): Stretched exponential kernel stochastic behaviour, $N = 1$; Top-right (b) Stretched exponential kernel mean-field behaviour; Bottom-left (c) Power law kernel stochastic behaviour, $N = 1$; Bottom-right (d) Power law kernel mean-field behaviour. The form of acceleration for fat-tailed kernels in the mean-field systems are in agreement with previous results [5, 23, 28], velocity increasing as a power law and exponentially with time for the stretched exponential and power law kernels respectively. For stochastic models, minimum replicates were: $L = 10^2$, 100; $L = 10^3$, 100; $L = 10^4$, 100; $L = 10^5$, 50; $L = 10^6$, 10; $L = 10^7$, 1.

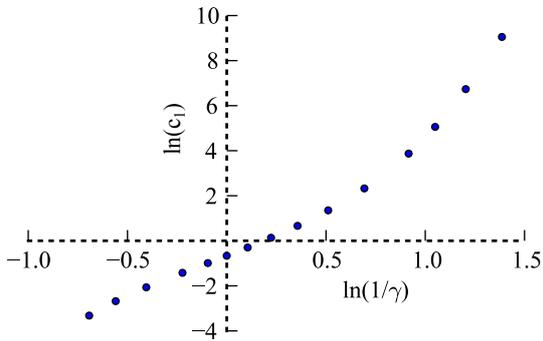

Figure 4: Logarithmic plot of the dependence of the front velocity on stretched exponential power $\gamma$. Note the apparent regime change close to $\gamma \approx 0.5$ ($\ln \frac{1}{\gamma} \approx 0.69$), rather than at $\gamma = 1.0$ as might be expected. The right hand side of the plot corresponds to long-range kernels with $\gamma < 1.0$. Plotting $c_1 = c - c_{\min \, [\text{Model 1}]} = c - 0.5$, rather than the actual velocity $c$, avoids a spurious saturation in the bottom left quadrant as $c_1 \to 0, \gamma \to \infty$.

When $\beta \leq 2.0$, our attempts to apply non-linear fitting to simulation results failed on two accounts. Firstly, the inferred prefactor to the velocity behaviour, $a_2$ in Table 1, becomes extremely small. Secondly, Akaike's information criterion now suggests that a stretched exponential fit is preferred, $L \approx e^{pT^q}$. It is therefore possible that power-law acceleration no longer adequately describes model behaviour when $\beta \leq 2.0$. However, as these waves travel extremely quickly it was not possible to confirm behaviour over long times.

The power law kernel with $\beta = 3.0$ represents the marginal case between the constant velocity and accelerated front regimes. Given this transition, we expect such invasions to display strong fluctuations, and indeed we observe a sharp peak in the relative standard error for filling time here. An example of filling behaviour in this region can be seen in Fig. 7.

**Bounds on the constant velocity of waves under stochastic Model 1, $N = 1$**

We can independently confirm the constant velocity results by obtaining estimates of the maximum and minimum wave velocities for each kernel. To do this, we perform a lattice version of the analysis conducted by Clark *et al* [37] for their dispersal model. We describe this in detail in Appendix 5. Briefly, the approach is as follows. In each generation, we consider the propagule dispersing furthest ahead of the wavefront as the "extreme disperser". This propagule defines the wavefront in the next generation, and the distance it travels indicates the wave velocity that generation. Given this, the maximum wave velocity can be estimated from the distribution of extreme dispersal distances when a large area of contiguous occupation stretches out behind the wavefront. Conversely, this distribution offers the minimum velocity when the wavefront consists of a single isolated occupied site. The asymptotic wave velocity results from our simulations lie between these maximum and minimum bounds, Fig. 5.

Our work indicates that Clark *et al* were correct in their prediction that dispersal kernels without a fat-tail lead to waves that are supported by a large region of occupation to their rear, thus attaining their maximum velocity. While fat-tailed kernels do lead to a more sparsely occupied wavefront, the minimum velocity was not achieved in our simulations. The implication is that this minimum value does not appear to be a good estimate of wave velocity given long-distance dispersal regimes. The importance of dispersal from behind the main wavefront is also supported by more complex models of plant dispersal [60].

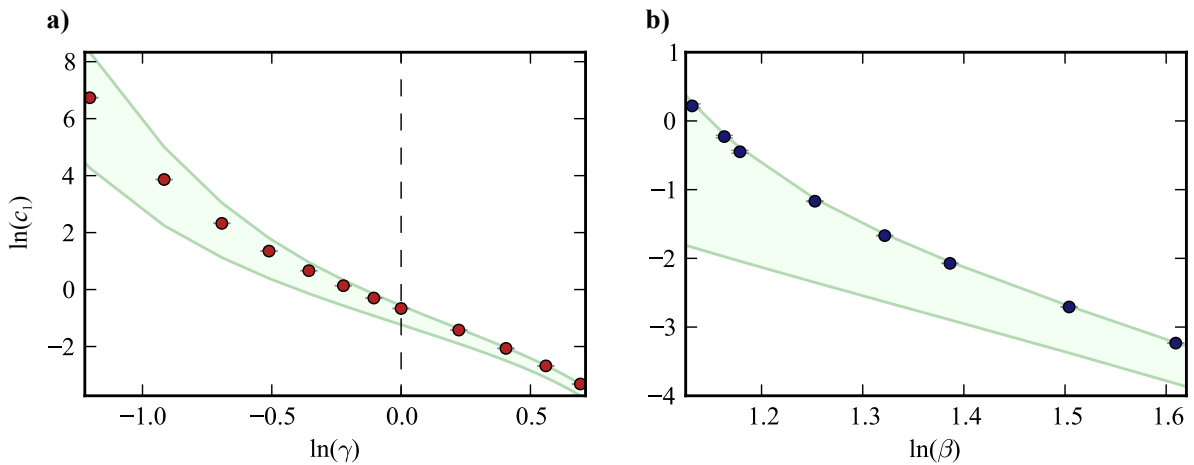

Figure 5: Placing bounds on velocity for the stochastic Model 1 simulations using a lattice variant of Clark *et al*'s method, see Appendix 5. Shaded area indicates the region between velocity minimum estimate with $N_s = 1$ and velocity maximum estimate with $N_s = 10^5$, where $N_s$ is the length of contiguous occupation behind the furthest forward site. Kernel length $l_{\max} = 10^5$, which is shorter than in our explicit simulations and may slightly bias upper velocity estimates downwards for the most fat-tailed kernels. a) Exponential family of dispersion kernels, $0.3 \geq \gamma \leq 2.0$, fat-tailed kernels are to the left; b) power law family of dispersion kernels, $\beta > 3$. Circles represent explicit Model 1, $N = 1$, simulation velocity results, with $L = 10^6$, minimum 10 replicates, based on time taken to reach 90% system filling. Other measures of velocity (eg. time until first dispersal to the right-most 10% of the system) give similar results, as do simulations with $L = 10^7$ (performed when $\gamma \leq 0.5$ or $\beta \leq 3.5$). Relative errors are shown, though were small.

*3.3. Model 2: Acceleration for fat-tailed kernel invasions in a mean-field system*

In our lattice mean-field simulations, dispersal according to a fat-tailed kernel leads to accelerating waves of advance. Importantly, this agrees with analytic predictions [5] for a spatially continuous version of the system. The signal of acceleration is apparent in the sub-linear scaling of filling time with system size (Figure 3b-d). Again, velocity behaviour obtained by non-linear regression is given in Table 1.

In the case of short-range kernels, $K(l) \propto e^{-|l|^\gamma}$ with $\gamma \geq 1.0$, the wave of advance travels at an asymptotically constant velocity. In the limit $\gamma \to \infty$, this case corresponds to the nearest-neighbour dispersal model, with $c = c_{\min \text{[Model 2]}} \approx 0.78$. For these kernels, we can independently verify simulation results. We do this by numerically obtaining the wave number and corresponding velocity according to eqs. (A3.7) and (A3.6) obtained in our marginal stability analysis of Model 2, Appendix 3. This approach yields extremely similar results to our full simulation studies, as shown in Fig.2.

When $\gamma < 1.0$, these kernels become fat-tailed, and acceleration is observed. The form of acceleration follows theoretical expectations quite closely (eg. [13, pp 176]), with filling behaving as $L \approx T^{\frac{1}{\gamma}}$ such that the acceleration $B \approx \gamma^{-1}$. When $\gamma = 0.5$, $B_1 = B - 1 \approx 1.0$, in agreement with the appropriately parameterised Eq.(21) of

[5].

Power law kernels, $K(l) \propto |l|^{-\beta}$, lead to waves of advance with extreme accelerating behaviour in the mean-field model. The form of acceleration is exponential, in agreement with previous studies [5, 23, 28], and can be modelled as a correction to random filling Eq. (15), using Eq. (14) $T \approx g \ln L + h$. We obtain a linear relation, $g \approx \frac{3}{2}\beta$ (see Table 1), reflecting the strong accelerating effect even when $\beta$ is large. Each order of magnitude increase in system size corresponds to a constant increase in filling time. If the continuous model is to be believed, dispersal kernels with this structure would have catastrophic implications in the case of a species invasion. No unusual behaviour is observed at $\beta < 3.0$, as might be expected given the divergence of the diffusion constant in this region.

Theoretical predictions suggests $g \approx \beta$ (eg. [13, pp 171-173]), such that while we recover the exponential form of acceleration, our simulations give a larger value of $g$.

### 3.4. Model 1, $N > 1$: Reducing dispersal stochasticity leads to slow filling time convergence with long-range kernels

As $N$ is increased, stochasticity due to dispersal decreases and we expect to approach the mean-field approximation of Model 2.

In Fig. 6 we present finite size scaling results comparing Model 1 with $N = 1, 10, 10^3, 10^5$ to the mean-field Model 2, for three fat-tailed kernels: the stretched exponential kernel with $\gamma = 0.5$ and the power law kernels with $\beta = 2.5, 3.5$. The two power laws kernels gave rise to different behaviour in the $N = 1$ stochastic system, but the same exponential acceleration in the mean-field case. The filling time for each $N$ relative to that of the mean-field model is also plotted. This shows the divergence between stochastic and mean-field models over time.

The stretched exponential kernel leads to an accelerating wave in the mean-field case. However, we find an asymptotically constant velocity wave in the highly stochastic $N = 1$ case, and even large $N$ systems rapidly converge to a constant velocity. For the power-law kernels, increasing $N$ leads to more extreme acceleration. When $\beta = 3.5$, transient acceleration persists for a longer period, but waves tend toward the constant velocity behaviour observed in the highly stochastic $N = 1$ system. Qualitatively, acceleration does not reach the extreme mean-field form of $T \approx g \ln L + h$ for either power law explored in our simulations, evident in the semi-log plot inserts in which a linear relationship is not achieved. The implication is that the degree of acceleration predicted for power law kernels using reaction-diffusion equations with fractional diffusion [23], for example, may not persist in real world populations due to stochastic effects.

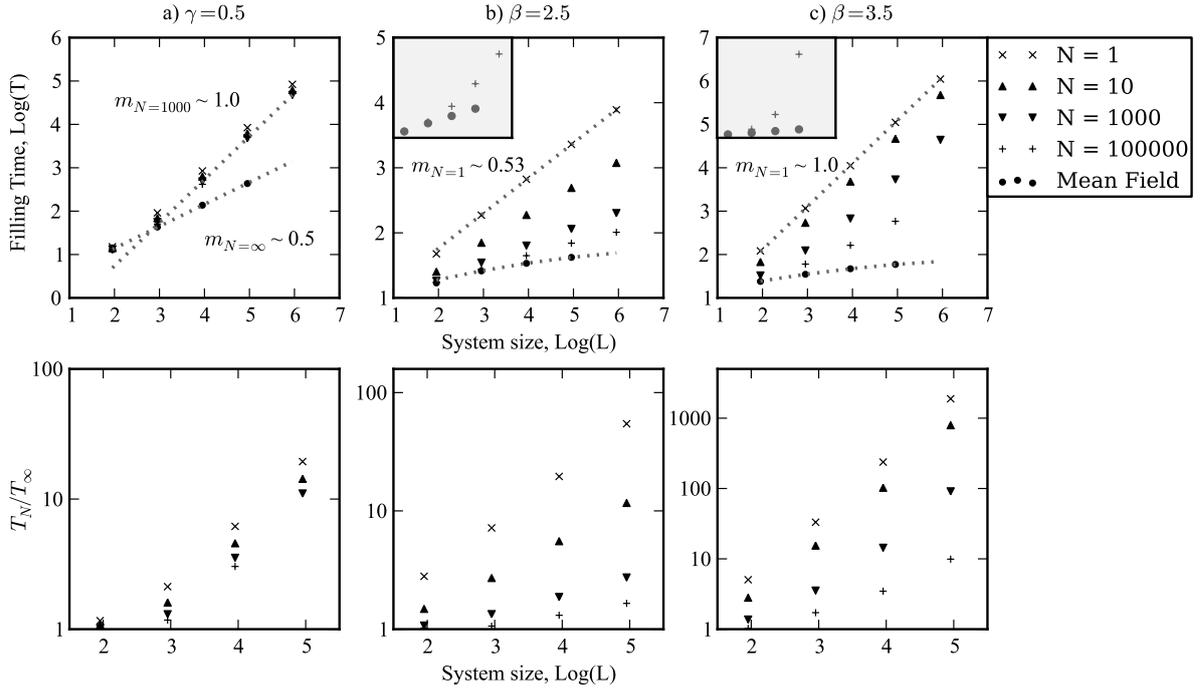

Figure 6: Comparing the behaviour of Model 1 with $N > 1$ to the mean-field prediction given 3 representative dispersion kernels. Upper plots show finite size scaling behaviour, lower plots the divergence of filling time predicted by each model. Plot a) follows a stretched exponential kernel, plots b) and c) power law kernels. Semi-log plots of $T$ against $\log L$ are inserted in the case of the power law kernels. Minimum replicates were: $10^2$: 100; $10^3$: 100; $10^4$: 50; $10^5$: 5; $10^6$: 1. Only a single simulation was conducted for the $\beta = 3.5, N = 10^5, L = 10^5$ system; otherwise, more simulations than the minimum number were possible. Relative error bars are shown in the upper plot, but are generally small.

The difference between stochastic and mean-field results for $T(L)$ increases with system size (see Fig. 6, lower plots). This reflects the greater wave acceleration apparent in mean-field systems, such that velocity diverges between the two models over time. We can use this as an indication of how the mean-field approximation deteriorates given the dispersal kernel and $N$. For small systems, or at early times, $L \approx 10^2$, the filling time results of stochastic and mean-field simulations are similar. For larger systems, eg. $L \geq 10^4$, results can diverge substantially. This is particularly severe for the shorter-range power law kernel ($\beta = 3.5$, $D \approx 1.15$) and the stretched exponential kernel ($\gamma = 0.5$, $D \approx 71.6$). Here, filling is $\approx 91$ and $\approx 11$ times slower respectively than the mean-field system when $N = 10^3$ and $L = 10^5$.

Figs. 7 and 8 show the effect of decreasing stochasticity on the structure of the wave of advance and on filling time for a $\beta = 3.0$ power law kernel and a stretched exponential with $\gamma = 0.5$. As $N$ increases, acceleration becomes more apparent at this scale for the power-law system but not for the stretched exponential - the former shows an increasingly concave interface between occupied and unoccupied space, and a substantial reduction in filling time. Acceleration occurs through jumps, with wave structure becoming patchier. This kernel is at the transition between well-defined and infinite variance, so rough and highly variable behaviour is expected on general grounds.

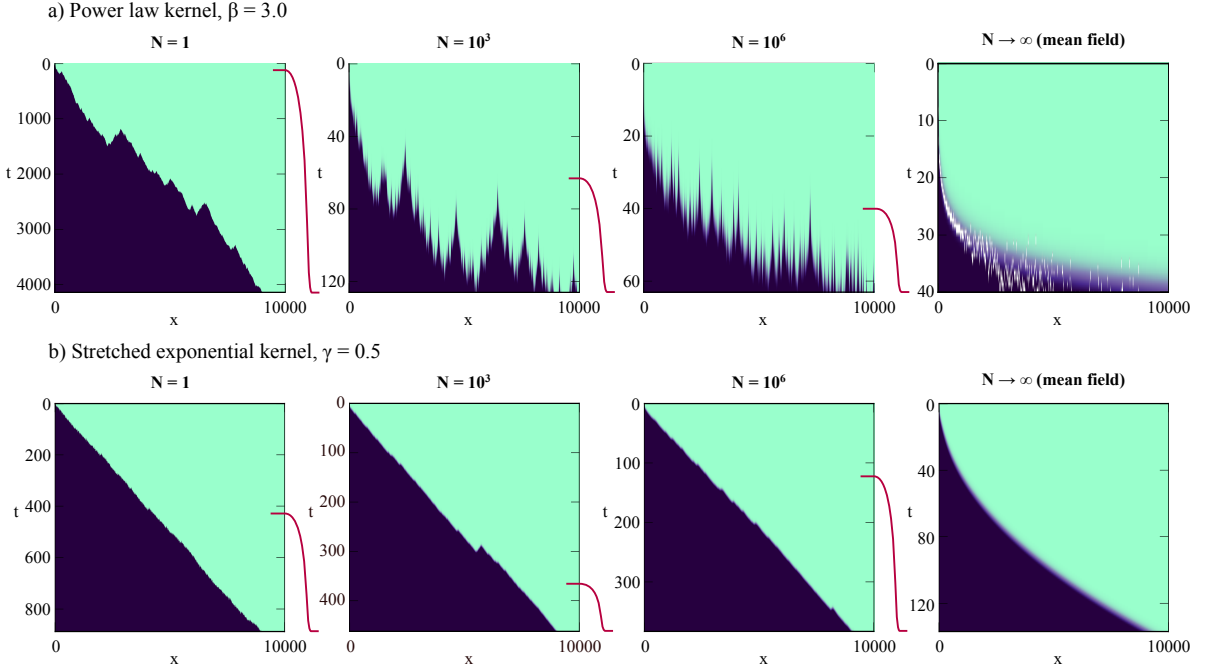

Figure 7: Plots of the filling processes of individual simulations using Model 1 with a range of $N$ values and two dispersion kernels. Shading represents population density at each position, $x$, on the 1-dimensional lattice (x-axis). Dark regions corresponds to complete square filling and light regions to empty space. Time progresses from top to bottom on the y-axis, such that systems begin with a single full site at the left-most position $x = 1$. Kernels are labelled accordingly; from left to right, $N = 1, 10^3, 10^6, \infty$ (mean-field). The time scale of each simulation has been rescaled to facilitate comparison of filling dynamics, with relative scale indicated to the right of each plot. The white shading on the upper mean-field plot with a power law kernel at $\beta = 3.0$ represents the filling process for $N = 10^8$, see text. System size, $L = 10^4$.

Regions of filling between 30% and 70% for the power law kernel with $N = 10^8$ have been shaded on top of the mean-field filling plot (top rightmost, Fig. 7a). Systems with extremely large $N$ approximate the mean-field dynamics closely at early times, but the patchiness created by stochasticity soon reappears.

The impact of increasing $N$ on wave of advance velocity has also been assessed for a short-range kernel (nearest-neighbour, see Fig. 2). Even with reasonably small $N$, wave velocity is close to mean-field predictions of $c \approx 0.78$ ($N = 100$, $c \approx 0.74$), suggesting that the mean-field approximation accurately reflects the underlying stochastic behaviour for very short-range kernels.

Our results indicate that the mean-field model gives a reasonable estimate of stochastic behaviour given short-range kernels, or for fat-tailed kernels at very short times, especially if $N$ is large. However, this and similar deterministic models, such as integrodifference equations, tend to over-estimate wave of advance velocity, an error that increases with time when long-distance dispersal is important. Even when filling time is well-predicted by the mean-field approximation, as in some very large $N$ systems, features such as the patchiness of the invasion are poorly described.

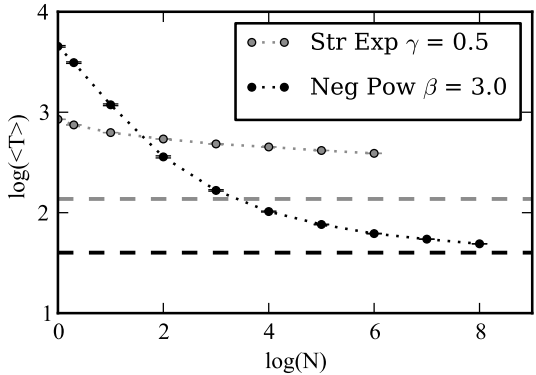

Figure 8: Convergence of filling time toward the mean-field limit (dashed line) as $N$ is increased for two fat-tailed kernels, $L = 10^4$. Replicates for $N = 1, 2, 10$ : 100; $N = 10^2$ to $10^5$ : 50; $N = 10^6$ : 10; $N = 10^7$ and $10^8$ : 1.

*3.5. Truncated power law kernels lead to asymptotically constant wave velocities*

Incorporating a long-distance cut-off to power law kernels has been found to describe well the movement patterns of various species [46–48], and has been investigated in the context of the Fisher-Kolmogorov equation with fractional diffusion [61]. Often, a limit to the distance at which dispersal can occur is also biologically reasonable, due to factors such as energetic constraints or a finite lifespan. It is also possible for such kernels to be retrieved from field data erroneously due to insufficient sampling of rare long-distance events, and an awareness of the errors that this might generate is useful. We here assess the impact of kernel truncation on wave velocity.

A basic prediction is that incorporating a cut-off by reducing $l_{\max}$, see Eq. 10, will lead to a slower wave of advance. A cut-off also causes power law kernels with $\beta < 3.0$ to have a defined variance. Asymptotically accelerating waves in either the mean-field or stochastic system are not expected.

We applied a long-distance truncation to power law kernels with various $\beta$, and to the stretched exponential case with $\gamma = 0.5$. Behaviour for the power law kernels is shown in Figure 9. In the mean-field system, the wave of advance accelerates according to the standard Model 2 until a time $t_0$. Velocity then approaches a finite value through a series of oscillations. The oscillations may be an example of the Gibbs phenomenon [62] due to the abrupt nature of our cut-off, and if so would not be expected in real systems.

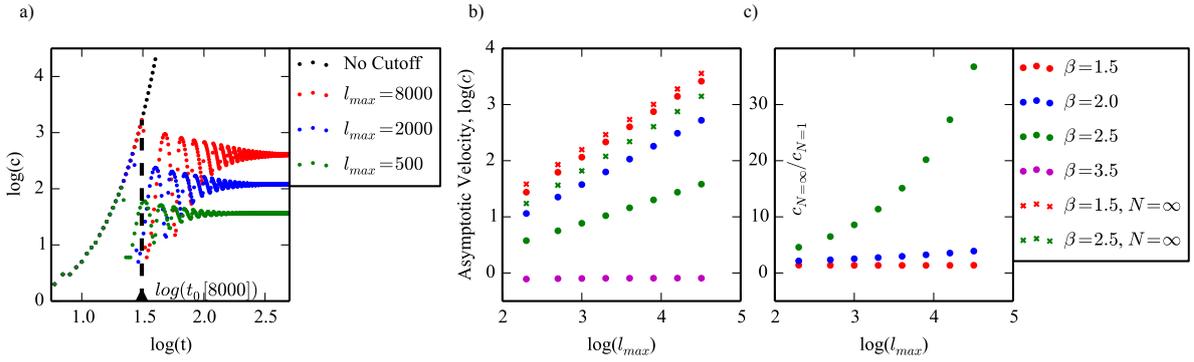

Figure 9: The impact of a long-distance cut-off on wave velocity given a truncated power law kernel. a) Log-log plot of velocity of wave front over time when $\beta = 2.5$, showing damped oscillations after a time period $t_0$. b) Log-log plot showing relationship between cutoff, $l_{\max}$ and asymptotic wave velocity, $c$, for both the stochastic ($N = 1$) and mean-field models. c) Relationship between asymptotic wave velocity reached in the mean-field system and the stochastic $N = 1$ system, for various power laws and sizes of cut-off. Stochastic systems were taken to have reached a stable velocity when the estimated velocity of the wave increased by no more than 1% upon an order of magnitude increase in system size. The stabilisation of mean-field systems was determined visually, with the $l_{\max} = 500$ case providing an example in a), velocity taken as the average filling increase over 10 generations after stabilisation. Replicates for stochastic systems were $L = 10^5$ : 100; $L = 10^6$: 50; $L = 10^7$: 20. Relative standard errors were minimal and are not shown

In the stochastic case, which might be considered a more realistic dispersal model, a finite velocity is again achieved for each dispersal kernel. Unsurprisingly, a cut-off will have particularly dramatic effects for kernels that would otherwise lead to accelerating waves of advance - power laws and stretched exponentials in the deterministic case, and power laws with $\beta < 3.0$ in the stochastic case. For power law kernels in the region $1.25 \leq \beta \leq 2.75$ with $N = 1$ we found that the heuristic fitting $c = a_3 l_{\max}^{(-0.26 \pm 0.09)(3-\beta)^2 + (0.97 \pm 0.18)(3-\beta) + (0.03 \pm 0.06)} \approx a_3 l_{\max}^{\frac{-(3-\beta)^2}{4} + (3-\beta)}$ captured asymptotic velocity behaviour. $a_3$ depends on $\beta$ and is $0.15 \leq a_3 \leq 0.7$ in this region, with the larger values observed when $\beta$ is closer to 3.

Asymptotic wave velocity in the mean-field Model 2 can also be fitted to $c = a l_{\max}^B$, but here $B$ remains close to 0.9 for power laws $1.25 \leq \beta \leq 3.5$ (see Fig. 9b) in the parameter space explored. However, as $l_{\max}$ becomes large there are indications of a gradual trend toward $B = 1$. This linear relationship is easily seen for the uniform kernel, which is recovered when $\beta = 0.0$. To further investigate this, we repeated the non-linear regression analysis, this time for $8000 \leq l_{\max} \leq 32000$ and $0.0 \leq \beta \leq 3.5$. The behaviour $c = (0.49 \pm 0.03)e^{-(0.99 \pm 0.08)\beta}l_{\max} + a_4 \approx 0.5 e^{-\beta} l_{\max}$ is supported.

Long-range cut-offs may apply to many biological kernels, and given this the asymptotically stable and substantially reduced difference between wave velocity in the stochastic and mean-field models is of some interest. The mean-field model particularly well approximates the stochastic case given low-$\beta$ power law kernels with moderately short-range truncation (Fig. 9c), and is also quite effective given a stretched exponential kernel with $\gamma = 0.5$ (where $\frac{c_{N=\infty}}{c_{N=1}} \approx 5.2$ when $l_{\max} = 5000$).

*3.6. Structural variations of our model support the generality of results*

As mentioned in the introduction, several studies have found that stochasticity breaks wave acceleration induced by Lévy flight dispersal kernels [37, 39]. Given that some of our simulations are seemingly at variance with these results, we have investigated three variations on the stochastic Model 1 presented above:

1. *Case 1* - Our stochastic Model 1 with $d > 0.0, b < 1.0$.
2. *Case 2* - A model following our Model 1 algorithm, but with the logistic effect applied to newborn individuals at the home site rather than the target site. The mean-field approximation is now Eq. (17). Note that this system does not result in an advancing wave when $N = 1$. For $N > 1$ and even, we set initial conditions to $n(0,0), n(1,0) = 0.5$.
3. *Case 3* - A stochastic version of Kot *et al*'s integrodifference model [5]. In each generation, a population growth stage occurs first within each site, such that every individual reproduces with probability $b[1 - \frac{n(x,t)}{N}]$ and dies with probability $d$. A dispersal stage then occurs, in which both newborn and older individuals disperse with probability $\Delta$. Usually, $\Delta = 1.0$.

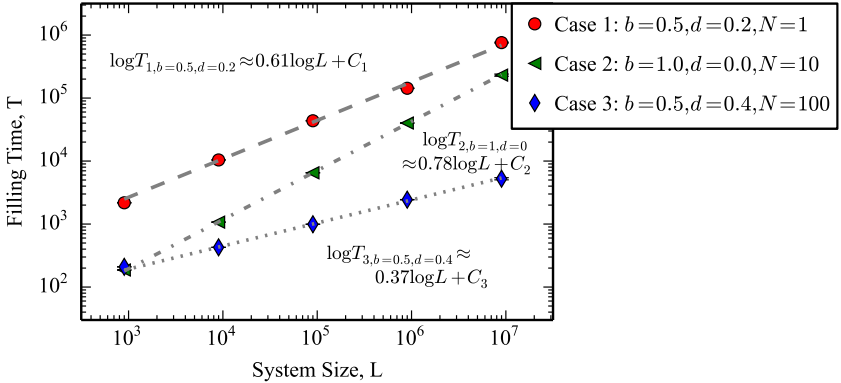

Figure 10: The impact of stochasticity on three variations of our Model 1, see main text. When dispersal follows a power law with $\beta = 2.5$, long-term acceleration of the wave of advance occurs in each case. Minimum replicates for different systems sizes were $L = 10^3, 10^4$:100; $L = 10^5$: 50; $L = 10^6$: 20; $L = 10^7$: 5. Relative standard errors are plotted, but are small.

In these models, we sometimes incorporate death, $d > 0$. An equilibrium system filling of $n(L) < 1.0$ is possible. We therefore chose to terminate the system when a site in the last 10% of lattice space has filling $n(x > 0.9L, t) \geq 0.1$. This approach is more closely allied with the traditional mathematical approach of tracking the furthest forward position with filling greater than some constant value, $n(x_\text{Max}, t) \geq n_\text{Min}$. We also tried terminating a Case 1 ($b = 0.5, d = 0.1$) system when system filling exceeds 50%, and found results to be qualitatively similar (not shown).

If the acceleration due to Lévy flights observed in Model 1 is quite general, we would expect it to hold under different implementations of birth and death, and of dispersal and crowding effects. Indeed, we find that acceleration is observed for the $\beta = 2.5$ power law kernel in each model variant (Fig. 10), and that velocity always increases as a power law with time. The different acceleration exponents, as indicated by the gradients of linear fittings in Fig. 10, are largely due to differing $N$. However, the structure of the model can also be important - with $N = 100$, the Case 2 system still accelerates comparatively slowly, as long as $d$ is small. We discuss this below. The three model variations also support results from Model 1 for the fat-tailed stretched exponential kernel, $K(l) = e^{-|l|^\gamma}$, $\gamma = 0.5$, which creates waves with asymptotically constant velocities (results not shown).

The specific implementation of the crowding effect has interesting implications. In Models 1 and 2, dispersal is logistically limited by occupation at the target site. In Case 2, the logistic effect is applied to new propagules from their home site. The equivalent mean-field equation for Case 2 is

$$n(x, t+1) = (1-d)n(x,t) + (1-d)b \sum_{l=-\infty}^{+\infty} K(|l|)n(x+l,t)[1 - n(x+l,t)]. \quad (17)$$

Simulations using this equation and the Model 2 mean-field equation, Eq. (6), lead to very similar results, even when long-range dispersal is important. The difference between filling times for a system of size $L = 10^5$ was less than 1% for stretched exponential kernels with $\gamma = 2.0, 1.0, 0.5$, and $\approx 2\%$ and $\approx 3\%$ for the power law kernels with $\beta = 2.5, 3.5$ respectively.

However, significant differences appear when stochasticity is introduced. Here, full sites frequently occur, but cannot contribute to population growth. Understandably, this effect is particularly pronounced when $d = 0$ and $N$ is small. For example, when $N = 4$ and dispersal follows a stretched exponential kernel ($\gamma = 0.5$), the Case 2 model leads to an asymptotic wave velocity of approximately $\dfrac{2}{3}$ that seen for Model 1. The effect is especially striking in the case of power law kernels, and the deviation between filling time results for Model 1 and Case 2 as $L$ increases is shown in Fig. 11a.

The impact of increasing death, $d > 0.0$, when the logistic effect arises at the home site is shown in Fig. 11b. Dispersal from sites that are far behind the main front and close to equilibrium filling is again possible, such that incorporating some death can increase wave velocity.

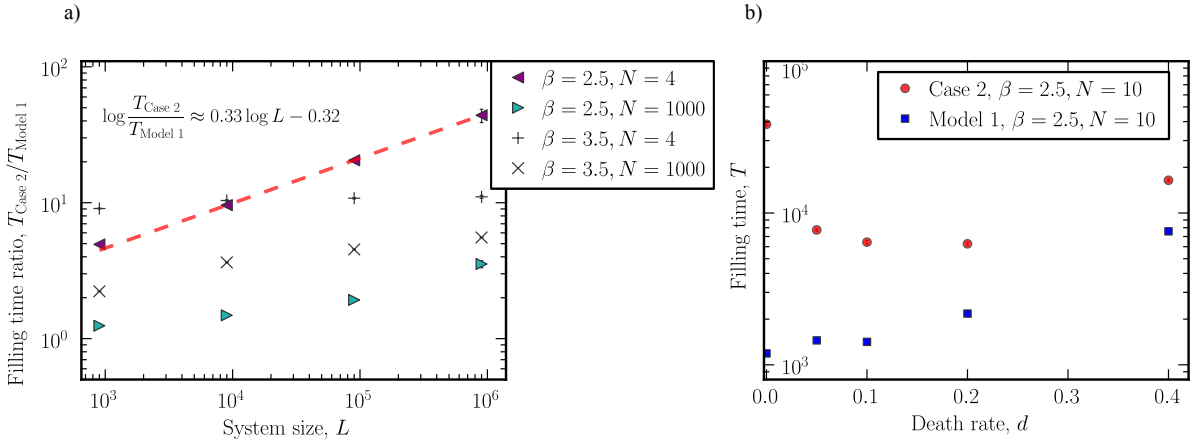

Figure 11: When only newborn individuals disperse in a stochastic system, wave velocity and acceleration depend on whether the logistic effect arises at the home site (Case 2, see text) or at the site targeted for dispersal (Model 1). This is particularly the case for low-$d$, small-$N$ systems incorporating long range dispersal. a) The ratio of Case 2 filling time to Model 1 filling time as system size $L$ is varied, with no death. Larger values indicate greater wave velocity deviations between the two systems. The ratio converges toward a constant value when the kernel leads to a finite-velocity wave, as for $\beta = 3.5$. b) Faster system filling occurs when $d > 0$ for the Case 2 system, $L = 10^6$. Minimum replicates are $L = 10^3 : 100$; $L = 10^4 : 50$; $L = 10^5 : 20$; $L = 10^6 : 5$.

## 4. Discussion

**Conclusions of our study**

When a population of an invasive species colonises a new region, it can grow and spread. This creates a wave of advance, which travels across the landscape with potential impacts on agriculture and natural ecosystems. Understanding the dynamics of invasions can help workers predict disruption and orchestrate a response. Simulation and mathematical models provide one route toward this understanding. In this study, we have identified important consequences of different modelling approaches, finding that stochasticity implied by a finite population fundamentally alters the behaviour of the wave of advance when long-distance dispersal is important. Results obtained by deterministic methods such as integrodifference equations or models of fractional diffusion risk significant inaccuracies in such circumstances, Figs. 6, 10.

Our simulations involve a simple model of population dispersal with tuneable demographic stochasticity. The models incorporate a flexible dispersal regime, with a logistic effect limiting the dispersal/growth process. In this way, the structure is similar to that of Mollison's simple epidemic [25], but occurs in discrete time, and indeed we confirm several of his fundamental results (for example, the different conditions on accelerating waves for stochastic and deterministic models). Some of these results appear to contradict, at first glance, recent work on stochastic models that incorporate long-range dispersal [37, 39]. We continue by discussing this discrepancy, after reviewing the core assumptions and detailed features of our simulation framework.

*Dimensionality*

Our models are one dimensional, a feature that is biologically plausible for the spread of species along rivers, coastal strips and island chains, but does not represent the general case. This approach is not unusual with most work on population spread historically conducted in one dimension for the sake of simplicity. Although for linear models the relationship between high-dimensional behaviour and the one-dimensional case is often straightforward, this is not true for non-linear stochastic models [54] such as ours. In the context of our work, there exists an obvious situation where dimensionality is important, specifically when dispersal occurs according to a power-law. In this case,

$$D = \int l^2 d^{d_I} l K(l) \sim \int_{r_{\min}}^{\lambda} l^2 l^{d_I - 1 - \beta} \mathrm{d}l = \int_{r_{\min}}^{\lambda} \frac{l^{d_I + 1}}{l^{\beta}} \mathrm{d}l \qquad (18)$$

with $d_I$ the dimensionality. We take a lower limit $r_{\min}$, defined so as to avoid divergence at small scales, and approximate the kernel using its power-law decay, which dominates as $\lambda \to \infty$. This integral describes the leading order behaviour of $D$ as a function of $\lambda$, and if it diverges $D$ does not exist. Given our results, stochastic Model 1 is expected to produce accelerating waves when $\beta \leq d_I + 2$.

To check whether the behaviour implicit in (18) is observed, we conducted limited simulations on a triangular lattice in two-dimensions. Our current implementation of the explicit simulation was impractical for lattices with sides of length greater than $10^4$, i.e. $10^8$ sites, in Model 1, and considerably smaller systems in the mean-field Model 2. Nevertheless, accelerating waves were seen in the mean-field system with fat-tailed dispersal kernels. This acceleration was found to break down when stochasticity was added, with the exception of systems with certain power-law kernels. We were unable to confidently determine the critical value of $\beta$ under which acceleration is preserved despite stochastic dispersal. For example, acceleration was not observed in the highly stochastic Model 1 system when $\beta = 6.0$, but was for the limited range of lattice sizes explored when $\beta = 5.0$. Equation (18) suggests that this is likely to be a transient feature.

An example of a similar system explored in two dimensions is the model of Kawasaki *et al* [40]. Here, the authors compare a stochastic cellular automaton model in which dispersal occurs randomly at a given rate with a deterministic variant. In both cases, site occupation is limited to 0 or 1. The deterministic case represents dispersal as a constant pressure exerted by occupied sites on empty sites within their dispersal range. Colonisation occurs when the cumulative dispersal pressure experienced by an empty site exceeds 1. The stochastic version of the model, which resembles our Model 1 with $N = 1$, yields a higher wave velocity than the deterministic case, even with the nearest-neighbour dispersal kernel. This is a result of the rougher wavefront for stochastic systems, such that there are more isolated sites, and the greater probability of propagules from isolated sites being successful.

Although this effect is in principle possible in our system, we did not observe greater wave speeds for our Model 1 than Model 2 in two dimensions. This is probably due to the potential for sites to have filling $0.0 \leq n(x,t) \leq 1.0$ in our deterministic model, such that newly arriving propagules immediately begin to contribute to the filling of nearby sites. There is a time-delay in the deterministic system of Kawasaki *et al*, as a site has to wait for sufficient dispersal pressure to transition from state 0 to 1. Despite this difference, the wavefronts in our stochastic Model 1 are rougher than those in the deterministic system. It is possible that both features are relevant, and a detailed characterisation of the behaviour of our model in two dimensions represents an important future step.

*Modelling assumptions*

A particular structural feature of our models is discreteness in time and space. Biologically, seasonal reproduction is seen in many species, and discrete-time models are often desirable. While one could consider discrete space to represent minimum territory size or regular fragmentation of a habitat, a more general interpretation views this as an artificially imposed lattice that averages effects within regions. This is a common approach used to simplify models, but can create an artefactual population crowding effect [63]. The lattice will also distort the dispersal kernel. Our replication of several forms of wave velocity behaviours caused by long-range dispersal [5, 23, 25, 28] suggests that our lattice dispersal kernels approximate sufficiently their continuous forms. Indeed, at long distances distortion to the dispersal kernel is minimal as the rate of tail decay is low.

A core subject of this paper is the relationship between deterministic and stochastic models of dispersal. We investigate the relationship between the stochastic Model 1 and a mean-field Model 2 by increasing the carrying capacity of sites, $N$. When $b = 1$ and $d = 0$, this only reduces the amount of stochasticity associated with the dispersal process. Model 2 is a mean-field model that offers a heuristically appealing approximation of Model 1, but we have not formally shown it to be correct. Indeed, in one case - that of stretched exponential kernels - the approach to mean-field wave velocity behaviour as $N$ is increased is particularly slow, Fig. 8. While this could indicate that another formulation is more appropriate, radically different behaviour in stochastic and mean-field systems for these dispersal kernels is theoretically expected [25].

Increasing $N$ in Model 1 also leads to a second transition, in that the filling impact of each propagule decreases. Biologically, this corresponds to the varied demographic dynamics of different species. Low $N$ systems resemble the dispersal of larger organisms, such as trees with few viable seeds per generation or various mammals. When $N$ is larger, organisms such as insects or micro-organisms are better described. Long-distance dispersal is possible for both, with the latter case often relying on aerial or aquatic currents, or living vectors (e.g. [64, 65]).

*Principle characteristics of model behaviour*

In this work, superficially similar models are found to yield qualitatively different conclusions. A fundamental difference in wave velocity behaviour is apparent between the stochastic and deterministic versions of our system. This sort of phenomenon has been noted by other workers in the context of long-range dispersal [25, 39]. The implication is that care is needed when designing models of species dispersal, and the consequences of mathematical or computational simplifications should be explored where possible.

Our simulations support the observation that fat-tailed kernels can give rise to accelerating waves of advance in deterministic models of population spread [5, 23, 25, 28]. The stochastic Model 1 also supports aspects of previous work [25]. Here, indefinite acceleration is seen to break down for some fat-tailed kernels. However, this is not always the case. Reflecting some previous results [25], but not others [37, 39], we find that kernels described by a power law with $\beta < 3$, which lack a finite second moment, can lead to long-term acceleration. Random walks according to similar distributions lead to Lévy flight superdiffusion. Our Model 1 is more complex, with a non-traditional branching rule and density-dependent dispersal. These differences do not appear to break the fundamental Lévy-flight-like dynamics. The apparent disagreement between this result and behaviour seen in similar systems [37, 39] is interesting, and we discuss it in detail below.

In Fig. 6 we examine the time evolution of the difference between our mean-field and stochastic systems. The mean-field Model 2 gives a reasonable estimate of the short-time wave velocity. However, at longer times the wave accelerates rapidly away from that predicted by the stochastic model. This effect is disrupted when a cut-off is applied to the kernel, §3.5, and in some cases (eg. low-$\beta$ power law kernels, see Fig. 9c) the velocities achieved for the mean-field and stochastic systems are rendered similar.

Acceleration observed in mean-field models seems to be caused by the logistic growth in low-occupation sites well ahead of the travelling wave. Demographic stochasticity usually disrupts this. However, when the dispersion kernel lacks a second moment, we regain acceleration. We speculate that this is a result of a superdiffusive effect from behind the main front. The advance of the wave is no longer strongly coupled with occupation close to the main front. This leads both to a patchy front and to a velocity that increases with total system occupation.

Should such conditions arise in the real world? In our introduction §1 we noted both that long-distance dispersal and accelerating waves of advance have been separately observed. Laying these points aside, we recall the extensive theoretical work on the Lévy flight foraging hypothesis [24]. When dispersal to new sites of reproduction is an extension of the foraging process, this hypothesis would suggest that the power law dispersion kernels are expected. Even when the spread of organisms to new territories is clearly separated from foraging, the potential to 'get lost' while foraging may facilitate rare long-distance dispersal events that would be expected to eventually dominate system behaviour. Anomalous kernels that resemble those explored above seem possible, although an eventual cut-off is inevitable in most real-world situations.

**Relationship to other work**

*The detailed form of reproduction*

Our observation of an accelerating wave of advance in a stochastic system due to power-law dispersal with $\beta < 3.0$ appears to contradict the results of Brockmann and Hufnagel [39]. In this study, the authors found that stochasticity disrupts Lévy flight superdiffusion in a two-particle reaction-dispersal system. Briefly, the model consists of a space containing a large number of particles of type $A$ and $B$. Both particles disperse by Lévy flights, and can also react according to $A_x + B_x \xrightarrow{k_1} 2A_x$ and $A_x + B_x \xrightarrow{k_2} 2B_x$. Our system is similar in having two particle states, 'empty' and 'full', and a transition of empty to full space due to population growth. However, in contrast, fluctuations do not tame superdiffusion caused by Lévy flights in our study.

We can suggest several reasons as to why this may be the case. The constant filling rate behaviour of the Brockmann-Hufnagel system is interpreted as being due to the probability of absorption outweighing local exponential growth when particles are rare. We see parallels between this and an emergent Allee effect. Given this, any critical difference between the models is likely to impact population growth in the low-population regime. The simple possibility that contrasting results relate to explicitly including stochasticity in birth and death is not supported, Fig. 10.

One fundamental difference between all our models and that of Brockmann and Hufnagel is that their system is Markovian, with reproduction and dispersal occurring independently and concurrently. The detailed manner in which dispersal and growth are combined in our system might explain the difference in behaviour. For example, an $A_x + B_x \to 2A_x$ reaction event in [39] leads to a local reduction in per-particle reaction rate, as in the Fisher-Kolmogorov equation, impacting both the parent and new-born organism. In our model, dispersal and growth are combined, such that a new propagule far ahead of the main front is likely to move to an unoccupied site. Neither parent nor new-born experiences a direct reduction in growth rate, although this is implicitly, and stochastically, implemented by local crowding. If such differences are critical, the implications for modelling long-range dispersal are quite profound, in that apparent subtleties of reproductive behaviour might lead to highly variable patterns of population spread. We certainly do not show this to be the case, however, and believe that further modelling or analysis of the two forms of system is necessary to clarify the importance of this effect.

Other explanations are plausible. The potential for both particles to disperse with Lévy flights in the Brockmann-Hufnagel system facilitates long-range counter-invasions. Alternatively, the acceleration in our systems may merely reflect an extremely long transient effect. This final point seems unlikely given the large systems explored and superdiffusive behaviour of Lévy flight random walks. Although acceleration sometimes appears to slow down slightly when $L \geq 10^7$, this is likely a consequence of our constructing the kernel to a finite distance $l_{\max} = 2 * 10^8$. Explicitly calculating the dispersal distance using the Hurwitz Zeta function appears to remove the effect.

*The extreme disperser approximation*

In the population dispersal literature, the model of Clark *et al* [37] resembles our simulations. The authors suggest that even dispersal kernels without a finite variance do not lead to indefinite acceleration, apparently at variance with our results (Figs. 3, 10).

The argument of ref. [37] focusses on 'extreme dispersers'. An occupied territory consists of one or more of organisms, each producing on average $R_0$ dispersing propagules per generation. The extreme disperser is the propagule that travels fur-thest ahead of the wavefront in a generation, and defines the wavefront in the next generation. Based on the increasing patchiness of occupied areas when dispersal is fat-tailed, the authors raise the possibility of a transition between two regimes. Initially, the wave of advance moves at the edge of a region of high population density. After a long period of time, the furthest forward individual in generation $t+1$ is merely the extreme disperser of the furthest forward individual at time $t$. The former case leads to the maximum wave velocity, while the latter approximates the minimum wave velocity. The authors suggest that this transition, combined with the fact that a sample of finite size from a kernel with infinite variance has finite variance, may lead such kernels to create constant velocity waves of advance.

We find that, for fat-tailed kernels with a finite variance, the minimum velocity approximation is not particularly effective in our system, §3.2. Furthermore, for Model 1 with $N=1$, our initial condition ($n(0,0)=1.0$) of a single isolated occupied site represents the minimum velocity case, and transient acceleration is nevertheless observed.

Clark *et al* [37] carry out explicit simulations of invasion dominated by extreme dispersal in the case of a bivariate Student's $t$ kernel [66]. This kernel is fat-tailed and lacks a well-defined second moment,

$$K(l) = \frac{1}{2\sqrt{2u}\left(1+\frac{l^2}{2u}\right)^{\frac{3}{2}}},\qquad(19)$$

where $u$ is a scale parameter for the distribution. This kernel closely resembles a power law at $\beta = 3$ over long distances, but unlike all our fat-tailed kernels it is convex at its source. Importantly, this power law represents the boundary case of undefined variance in one dimension, and we expect acceleration caused by such kernels to be marginal and difficult to detect. Indeed, we have performed simulations using the bivariate Student's $t$ kernel to assess the importance of convexity at source, and results closely reflect the power law at $\beta = 3$. Specifically, the wave accelerates for some time, but appears to approach a constant velocity eventually (for details, see Appendix 6). The precise from of the short-range dispersal regime does impact the length of transient acceleration, but does not change the long-term dynamics. For example, the time taken for acceleration to cease is greater when parameter $u$ is large, as is the case for the species of spruce (*Picea*) modelled by Clark *et al*, where $u = 5531$.

We therefore suspect that the constant velocity behaviour suggested by Clark *et al* for the bivariate Student's $t$ kernel reflects the fact that it is the boundary case for accelerating waves (i.e. $\lim_{l\to\infty} K(l) \propto |l|^{-3}$), rather than the finite number of dispersal events per generation or factors such as its short-range behaviour. Acceleration may nevertheless be visible when using different methods of data analysis. We expect that further work will resolve behaviour caused by such marginal dispersal kernels.

Our work suggests that the method used by Clark *et al* would give unreliable results for kernels with an undefined second moment that are non-marginal, as these are expected to accelerate due to the dispersal impact of sites far behind the main front. However, we note that the framework provides very accurate estimates of velocity behaviour for waves generated by power law kernels with $\beta > 3$, Fig. 5, using the maximum rather than minimum velocity approximation. The method is also reasonably effective for stretched exponential kernels when $\gamma$ is not too far below 1.0, and it therefore remains very interesting in these two cases. It is also certainly possible that further refinement to the approach will allow the recovery of the velocity given by low-$\gamma$ stretched exponential kernels, or the acceleration behaviour caused by power law kernels with $\beta < 3.0$.

**Extending our model**

As discussed in §2, our method of combining birth and dispersal leads to significant differences from the traditional Fisher-Kolmogorov equation. Whether one interprets our approach as the release of dispersing propagules, or as the dispersal of adults after their generation-long maturation and local reproduction, our algorithm doesn't accurately represent the behaviour of certain organisms. Given this, we performed several extensions to our model, §3.6.

These modifications consisted of incorporating stochasticity in birth and death ($b > 0$, $d < 1$), applying the logistic effect at the home site as in the Fisher-Kolmogorov equation, and implementing a stochastic version of the integrodifference equation studied by Kot *et al* [5], Eq. (5). In each case, acceleration of the wave of advance was observed in the stochastic system given Lévy flight dispersal. These variations and re-interpretations of our model suggest our results are, qualitatively, quite general.

The detailed implementation of the logistic effect has interesting modelling implications. When only propagules disperse and these are subject to intra-specific competition, applying the logistic effect at the home site (Case 2 in §3.6), as is traditional in population dispersal modelling [5, 16], can cause the wave of advance to stop before complete filling. This occurs when all occupied sites are completely full, and is inevitable when $N = 1$. If $N > 1$, the wave of advance should not stop permanently unless $d = 0$. When $d > 0$, transient pauses are still possible, and are more likely when $N$ is small, dispersal is short-range and $d \ll 1.0$.

With respect to long-range power-law dispersal ($\beta < 3.0$), the rate of acceleration tends to be substantially lower for the Case 2 model than in Model 1. The significant influence on the wave of advance from sites behind the main front is damped, and we might expect velocity to scale with the total number of partially-full sites. For this reason, death has an unusual effect in such systems, and the maximum wave speed is sometimes achieved when $d > 0$, Fig. 11. This behaviour is interesting from a modelling perspective, and although the Case 2 system is somewhat contrived (particularly when $d = 0$) the effect may in principle play a role in real species invasions. Relevant situations include cases in which crowding has a significant impact on parental reproductive potential, or where newborn offspring experience intense intra-specific competition before dispersal from the home site. Slight changes to the stochastic algorithm, such as allowing reproductive individuals to disperse, are likely to have a significant impact on the behaviour of this model.

**Final Remarks**

Taken together, our results suggest that the modelling strategy employed should depend strongly on the dispersal regime under consideration, the time-scale of interest, and the life-history details of the organism in question. In general, methods that poorly represent the long-distance tail of dispersal regimes, such as integrodifference equations, should be used with caution when these regions determine wave behaviour. Such methods are frequently encountered in the literature, and are often applied specifically for their ability to include long-range dispersal kernels. There is a danger here of using sophisticated mathematics to derive qualitatively incorrect conclusions. Our work helps to clarify the conditions under which random effects due to demographic stochasticity are important, and the severity of errors expected when they are ignored.

Comparing our results to other work highlights the potential for subtleties in model design to create apparently contradictory system behaviours. This does not necessarily reduce the validity of the different approaches. However, which method is appropriate will depend critically on the real-world scenario one is seeking to explore. Conclusions for one field or problem may not translate simply to other applications; and, perhaps worryingly, it can take structural investigation of a model rather than basic parameter sweeps to identify this.

**Acknowledgements**

We would like to thank Lee Hazelwood (University of Cambridge), Anne Kandler (City University) and James Steele (UCL) for important input at the beginning of this project. We also appreciate helpful comments from Dirk Brockmann (Humbolt University), and from Tim Reluga (Penn State) and two anonymous reviewers.

# Supplementary Information

**Long-range kernels, stochasticity and the broken accelerating wave of advance**
G. S. Jacobs and T. J. Sluckin

## A1. Simple derivation of a diffusive limit for mean-field Model 2

We here offer a basic derivation of a diffusive limit for our mean field Model 2. This is intended to highlight differences between our model and the Fisher-Kolmogorov equation, rather than as a rigorous indication that this is the exact diffusive limit.

We begin with our Model 2 equation,

$$n(x,t+1) = (1-d)n(x,t) + (1-d)b[1-n(x,t)] \sum_{l=-\infty}^{+\infty} K(|l|)n(x+l,t), \quad \text{(A1.1)}$$

noting that in the temporal continuum limit

$$[n(x,t+1) - n(x,t)] \to \frac{\partial n(x)}{\partial t}, \quad \text{(A1.2)}$$

leading, with $d=0$ and $b=1$, to the integrodifference equation

$$\frac{\partial n(x,t)}{\partial t} = [1-n(x,t)] \sum_{l=-\infty}^{+\infty} K(|l|)n(x+l,t). \quad \text{(A1.3)}$$

We take the lattice sum on the right hand side to its spatially continuous limit,

$$\sum_{l \neq 0} K(|l|)n(x+l,t) \to \int_{-\infty}^{\infty} K(|x-y|)n(y,t)\mathrm{d}y, \quad \text{(A1.4)}$$

where the normalisation conditions

$$\sum_{l=-\infty}^{\infty} K(|l|) = \int_{-\infty}^{\infty} K(|x-y|)\mathrm{d}y = 1. \quad \text{(A1.5)}$$

ensure that the kernels represent probabilities of propagules at particular points.

We now suppose the integral to be one dimensional and follow the traditional approach (eg. [67] in a rather similar model of epidemic spread) of expanding $n(x')$ in a Taylor series around $x$ and substituting $n(x')$ into Eq. (A1.4), omitting forgotten terms. A similar method can be applied to integrodifference equations with the non-linear behaviour applied to the growth term, and retrieves the Fisher-Kolmogorov equation (eg. [27]).

Some caution is advised here - [27], following [26], find that complications can arise when describing the discrete-time behaviour as a continuous-time system, particularly when a significant time-delay is involved. This is relevant for many biological systems. However, as our intention is to present a qualitative comparison to the Fisher-Kolmogorov model rather than to retrieve a diffusive approximation for further analysis we trust this derivation will suffice. We obtain

$$n(x') = n(x) + (x' - x)\frac{\partial n(x)}{\partial x} + \frac{1}{2}(x - x')^2\frac{\partial^2 n(x)}{\partial x^2} + \dots. \quad \text{(A1.6)}$$

and hence

$$\frac{\partial n(x,t)}{\partial t} = n(x,t)(1 - n(x,t)) + D(1 - n(x,t))\frac{\partial^2 n(x,t)}{\partial x^2}, \quad \text{(A1.7)}$$

with

$$D = \frac{1}{2}\left(\int_{-\infty}^{\infty} l^2 K(|l|)\,\mathrm{d}l\right),$$

which is the diffusion approximation of Mollison's simple epidemic [52]. Indeed, we find that several analytic results for this model [52] hold when we reduce the spatial and temporal scale of our system, §3.1.

This equation resembles the Fisher-Kolmogorov equation, though is not identical to it. Although generally our simulations have $d = 0$ and $b = 1$, we do explore the situation where $d > 0$, and re-introducing both terms modifies Eq. (A1.7) to yield

$$\frac{\partial n(x,t)}{\partial t} = \alpha n(x,t)\left(1 - \frac{n(x,t)}{\kappa}\right) + \tilde{D}[1 - n(x,t)]\frac{\partial^2 n(x,t)}{\partial x^2}. \quad \text{(A1.8)}$$

where the equilibrium occupation is given by $\kappa = \dfrac{\alpha}{(1-d)b}$, the Malthusian constant is $\alpha = (1-d)b - d$, and the effective diffusion constant is also modified, yielding $\tilde{D} = b(1-d)D$.

On inspection, our diffusive approximation Eq. (A1.8) behaves in a similar manner to the stochastic algorithm of Model 1 and our mean-field equation Eq. (6). As a result of the detailed form of reproduction in our model, and in contrast to the Fisher-Kolmogorov equation, diffusion neither occurs when $b = 0$ or when the system is full. In the latter case, increasing the death rate $d$ leads to a carrying capacity $\kappa < 1.0$, and diffusion is resumed. The population dies out when $d > (1-d)b$, as the number deaths a site experiences outweighs its contribution to population growth through surviving births, even in an empty environment. These behaviours are biologically reasonable in many cases. For example, the main crowding effects for certain plant species are likely to arise through limiting the survival of recently dispersed saplings rather than in reducing the number of propagules for adult organisms.

We discuss the relevance of these differences given the Linear Conjecture in the main text. Results from simulations that investigate the effect of implementing a logistic effect based on home-site filling can be found in §3.6.

### A2. Periodic boundary conditions

How one treats boundary conditions is a methodological question encountered in many modelling scenarios. Frequently, periodic boundary conditions are applied, which can make a system 'neater' from a physical perspective. However, in explicit simulation particularly, we have flexibility in this regard, and conditions that greater resemble the scenario being modelled are appropriate. We have chosen to ignore dispersal outside the system in the majority of our simulations, which is reasonable if we are modelling invasion of a stretch of viable habitat along a coastline or river. Periodic boundary conditions better resemble the coast of an island for which the entirety of its coast is habitable for a species, but dispersal cannot occur over the main landmass.

To consider this case, and for modelling completeness, we here briefly present results for the highly stochastic Model 1 system with periodic boundary conditions. Our essential results for the Model 1 stochastic system are preserved in the case of periodic boundary conditions, with wave acceleration only observed for power law kernels with $\beta < 3.0$.

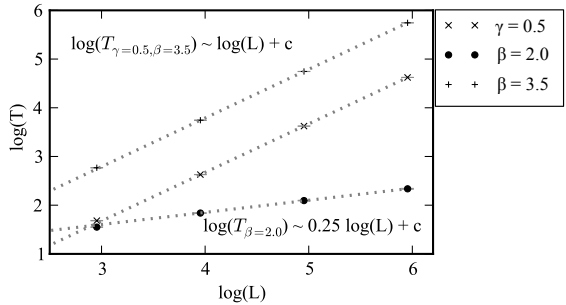

Figure A1: Filling time behaviour for Model 1 with periodic boundary conditions. Constant velocity is suggested for waves arising from both the stretched exponential kernel, $\gamma = 0.5$, and the power law kernel with $\beta = 3.5$. Acceleration is apparent for the power law kernel with $\beta = 2.0$, as evident in the log-log gradient of filling time against system size $< 1.0$. The acceleration parameter of 0.25 is estimated as similar to that of our main simulations ($\sim 0.22$).

Under periodic boundary conditions, long-range dispersal at distances far greater than the system size roughly equates to random positioning of the propagule after a large number of 'laps' of the lattice, creating the possibility of unrealistic dispersal events for smaller system sizes.

### A3. Marginal stability analysis of Model 2 with a nearest-neighbour kernel

Here we offer an argument based on standard methods to predict the velocity implied by our mean-field Model 2 for short-range kernels. The full recurrence relation of Model 2 is given by Eq.(6). Far ahead of the wave of advance, all $\{n\}$ are small, and we can linearise, yielding:

$$n(x, t+1) = n(x,t) + b \sum_{y \neq x} K_{xy}\, n(y,t). \tag{A3.1}$$

If the kernel is not fat-tailed, we can parameterise the asymptotic behaviour as:

$$n \approx e^{-(kx-\omega t)} \tag{A3.2}$$

where $k$ is unknown. Eq.(A3.1) yields a dispersion relation $\omega(k)$. It is a standard result that the dominant $k = k_c$ is given by the minimum value of $c(k) = \dfrac{\omega(k)}{k}$ [55, 57], where $v = c(k_c)$ and $c(k)$ is the speed of the wave with wave number $k$. For general kernels, substituting into Eq.(A3.1) yields:

$$e^{-[kx-\omega(t+1)]} = e^{-(kx-\omega t)} + b \sum_{y \neq x} K_{xy} e^{-(ky-\omega t)}. \tag{A3.3}$$

and hence, given the symmetry of $K_{xy} = K_{yx} = K(|x-y|)$,

$$e^{\omega} = 1 + b \sum_{y \neq x} K(|y-x|) e^{k(x-y)} \tag{A3.4}$$

$$= 1 + b \sum_{n \neq 0} K(|n|) e^{kn}. \tag{A3.5}$$

Taking logarithms and then applying the definition of the dispersion relation yields an expression for the velocity as a function of wave number.

$$c(k) = \frac{1}{k} \ln\left(1 + b \sum_{n \neq 0} K(n) e^{kn}\right). \tag{A3.6}$$

Differentiating (A3.4) in terms of $k$ then implies

$$\frac{d}{dk}(ck) e^{ck_c} = b \sum_{n \neq 0} n K(n) e^{k_c n}$$

$$\frac{\omega}{k_c} = \frac{b \sum_{n \neq 0} n K(n) e^{k_c n}}{1 + b \sum_{n \neq 0} K(n) e^{kn}}$$

$$\ln\left(1 + b \sum_{n \neq 0} K(n) e^{k_c n}\right) = \frac{b \sum_{n \neq 0} (n k_c) K(n) e^{k_c n}}{1 + b \sum_{n \neq 0} K(n) e^{k_c n}}. \tag{A3.7}$$

The value of $k_c$ can be determined using Eq. (A3.7), and $c = c(k_c)$ is obtained by substituting this into Eq. (A3.6). If $b < 1$, one can consider this a change in the time scale by normalising the resulting velocity by a factor $\frac{1}{b}$.

In the nearest neighbour kernel case, Eq. (A3.4) reduces to

$$e^{\omega(k)} = 1 + \cosh k. \tag{A3.8}$$

Substituting Eq. (A3.8) into Eq.(A3.7), we obtain

$$\ln\left(1 + \cosh k_c\right) = \frac{k_c \sinh k_c}{1 + \cosh k_c},$$

with

$$c(k) = \frac{\ln\left(1 + \cosh k_c\right)}{k_c} \tag{A3.9}$$

following Eq.(A3.6). Solving this equation numerically also yields $c(k_c) = v = 0.78$, which serves as a important check on the accuracy of the simulation results.

### A4. Clarifying kernel behaviour

The diffusion constant, $D$, is the traditional variable used to describe dispersal in the Fisher-Kolmogorov equation Eq. (1). This value was determined for various kernel parameterisations using the root mean square displacement of five thousand 50,000-step random walks and Eq. (2). The accepted $D$ was the average of 50 repetitions of this process. Velocity for constant speed systems was retrieved from the filling time of systems with $L = 10^6$, using Eq. (11).

Having obtained values for $D$ and $c$ for different kernel parameters, we can estimate the relationship between these quantities using non-linear regression, Fig. A2. This identifies slight departures from the $c \approx 2\sqrt{\alpha D}$ relationship, Eq. (4), derived from the Fisher-Kolmogorov equation. Given that a diffusion approximation of our model shows important differences from the Fisher-Kolmogorov equation, see Appendix 1, this is not surprising. A better point of comparison is the simple epidemic explored by Mollison [25, 52], which is very similar to our model but is continuous in space and time, and again does not adhere to Eq. (4). Indeed, by increasing the spatial and temporal resolution of our model, we are able to retrieve velocity results derived for the simple epidemic. The implication is that the relationships we find between $D$ and $c$ reflect both the structure of our model and discretisation effects.

We approximate $c$ for a continuous-time, continuous-space model by applying the linearisation method detailed in Appendix 3. This time, $b$ is reduced to $10^{-5}$ so as to substantially increase the temporal resolution of the system. The spatial detail is increased by using the modified kernel $K(\frac{l}{\phi}) = e^{-|\frac{l}{\phi}|^\gamma}$, $l \in \mathbb{Z}$, with $\phi = 10^5$. The value of $D$ in continuous space is approximated using this kernel and the method described above. For the deterministic approximation of the simple epidemic, an exponential kernel is known to lead to a wave of advance with $c \approx \frac{3\sqrt{3}}{2} D^{\frac{1}{2}}$, while normal kernels give $c \approx \sqrt{2De}$ [52]. These results are retrieved almost exactly, Table A1. Finally, we can explore the behaviour of $c$ given $D$ for our exponential family of kernels, $\gamma \geq 1$, by applying non-linear regression on our results. We identify the relationship $c \approx 2.5992 D^{0.5777} \approx \frac{3\sqrt{3}}{2} D^{\frac{1}{\sqrt{3}}}$ for this approximation of a continuous-space, continuous-time system.

| Time scale ($b$) | | 1 | $10^{-5}$ | 1 | $10^{-5}$ | $D$ | | Wave velocity estimate | |
| --- | --- | --- | --- | --- | --- | --- | --- | --- | --- |
| Lattice scale ($\frac{1}{\varphi}$) | | 1 | 1 | $10^{-5}$ | $10^{-5}$ | Discrete | Continuous | FKPP | Simple Epidemic |
| $\gamma =$ | 2, | 0.875 | 1.645 | 0.654 | **1.166** | 0.572 | 0.25 | 1 | **1.165** |
| | 1.75, | 0.946 | 1.757 | 0.721 | 1.276 | 0.638 | 0.290 | 1.078 | |
| | 1.5, | 1.064 | 1.948 | 0.833 | 1.458 | 0.763 | 0.370 | 1.217 | |
| | 1.25, | 1.288 | 2.309 | 1.045 | 1.802 | 1.019 | 0.530 | 1.457 | |
| | 1.1, | 1.538 | 2.710 | 1.283 | 2.185 | 1.335 | 0.739 | 1.719 | |
| | 1.0 | 1.809 | 3.142 | 1.544 | **2.598** | 1.715 | 1 | 2 | **2.598** |
| $c = \mu D^\rho$, | $\mu$ | 1.2684 | 2.2859 | 1.5452 | 2.5992 | | | | |
| | $\rho$ | 0.6565 | 0.5891 | 0.6184 | 0.5777 | | | | |

Table A1: Relationship between wave velocity and diffusion constant for mean-field Model 2 given a stretched exponential dispersal kernel, $K(\frac{l}{\varphi}) = e^{-|\frac{l}{\varphi}|^\gamma}$, $l \in \mathbb{Z}$. The spatial scale is given by $\frac{1}{\varphi}$ and temporal scale incorporated as the birth rate, $b$. Low values correspond to a fine scale in both cases. Death is ignored in each case, such that $b$ is essentially the traditional Malthusian parameter, $\alpha = b - d$. $D$ and $c$ were estimated as described in the main text. Velocity predictions derived from the Fisher-Kolmogorov equation follow the formula $c = 2\sqrt{\alpha D}$, while those according to the simple epidemic use formula given in [52]. The dependence of $c$ on $D$ was estimated for each system by non-linear regression on log-transformed data. When $b, \varphi = 1$ this relationship corresponds well with that obtained through explicit simulations, Fig. A2.

We have also checked the accuracy of $D$, as estimated through the random walk simulations. This involved obtaining the values of the diffusion constant for power laws with $\beta > 3.0$ by applying Eq. (3), yielding the formal relation:

$$D_\beta = \frac{I_{\beta-2}}{2I_\beta}, \tag{A4.1}$$

where

$$I_\beta = \sum_{n=1}^{\infty} \frac{1}{x^\beta} = \zeta(\beta) \tag{A4.2}$$

is the Riemann zeta function. The close correspondence between predicted and observed values of $D$ shown in Fig. A3 serves as a simple check on our random walk results. Note the divergence of the sum at $\beta - 2 = 1$, corresponding to the loss of a finite second moment at this point.

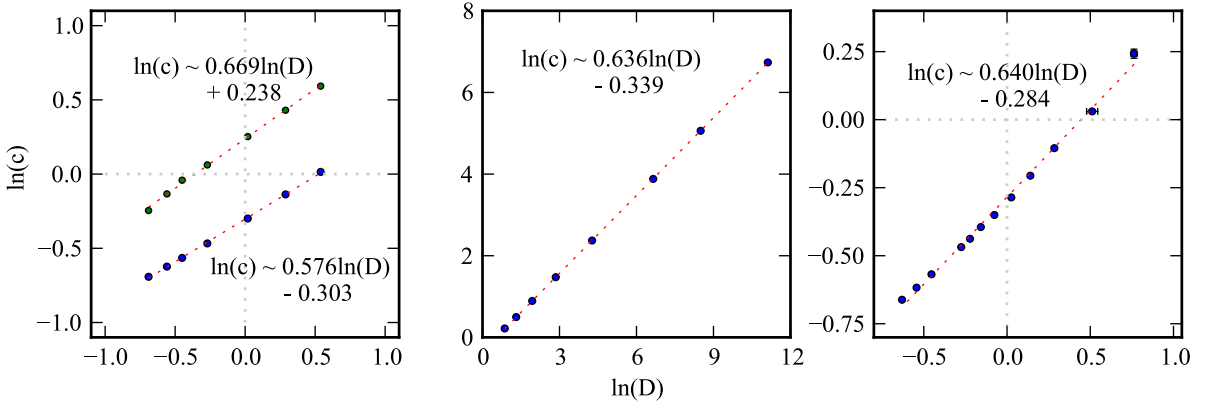

Figure A2: Relationships between diffusivity, $D$, and asymptotically constant wave velocity, $c$, under various dispersal kernels. For stochastic systems, $N = 1$. Left - exponential family kernels, $\gamma \geq 1.0$, stochastic Model 1 (blue) and mean-field Model 2 (green); centre - stretched exponential kernels, $\gamma < 1.0$, stochastic Model 1; right - power law kernels, $\beta \geq 3.0$, stochastic Model 1.

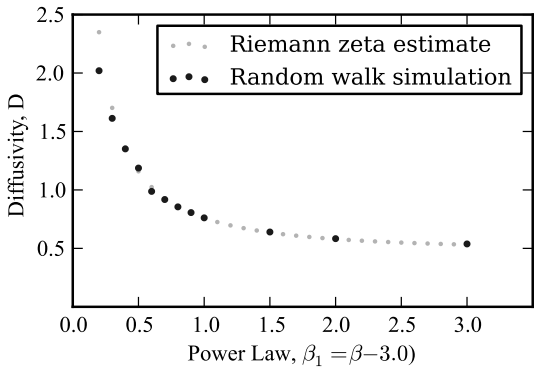

Figure A3: Diffusion constant for power law kernels, estimated both numerically and by simulation of 5,000 random walks to time 50,000. Results show close correspondence, but begin to diverge as $\beta$ approaches 3.0.

**Estimating $D$ for power law kernels when $\beta \gtrsim 3.0$**

We can estimate the value of $D$, where it exists, for power law kernels in the region $\beta \gtrsim 3.0$ as follows. In general,

$$K_\beta(l) = \frac{|l|^{-\beta}}{\sum_{l=1}^{\infty} l^{-\beta}} = \frac{1}{l^\beta \zeta(\beta)}, \qquad (A4.3)$$

where $\zeta(\beta)$, as above the Riemann zeta function, is the correct normalisation factor. In the region $\beta \gtrsim 3.0$, the principal contributions to the diffusion constant come from very large $l$. The lower limits (close to zero) are unimportant, and the sum can be replaced by an integral, as in Eq. (3). We truncate the lower limits in order to avoid the unphysical low $l$ divergence. Then the diffusion constant is given by Eq. (3), and is here

$$D_\beta \approx \frac{1}{2\zeta(\beta)} \int_1^\infty \frac{l^2}{l^\beta} dl. \qquad (A4.4)$$

The key part of this integral is in the integral of $\frac{1}{l^{(\beta-2)}}$, which diverges when $\beta - 2 = 1$,

or $\beta = 3$. Close to $\beta = 3$ the divergence dominates the behaviour:

$$D_\beta \approx \frac{1}{2\zeta(3)} \frac{1}{(\beta - 3)}. \tag{A4.5}$$

Once $\beta \leq 3.0$ the integral diverges, and $D$ is no longer defined.

**Behaviour of $D$ in stretched exponential kernels**

We can also consider the behaviour of $D$ for the stretched exponential kernels, $K(l) \propto x^{-|l|^\gamma}$. We obtain the variance of a spatially continuous stretched exponential kernel by applying the expression

$$\theta^2 = \frac{\Gamma\left(\frac{3}{\gamma}\right)}{\Gamma\left(\frac{1}{\gamma}\right)}, \tag{A4.6}$$

given in [58]. We can retrieve the diffusion constant using $D = 0.5\theta^2$, and obtain values that closely correspond to our random walk procedure on a fine spatial lattice ($\phi = 10^{-5}$), presented in Table A1. For the spatially discrete system, there are, unsurprisingly, deviations. Using (A4.6) to estimate $D$ is generally better for moderately small $\gamma$, with the difference less than 20% when $0.3 \leq \gamma \leq 0.5$; for $\gamma = 0.9$ the difference is about 60%. This is related to the growing role of long-range dispersal as $\gamma \to 0.0$, such that the long tail, where spatial discretisation has least impact, increasingly dominates dispersal behaviour.

We can use these values to roughly estimate $c$, either by assuming a Fisher-Kolmogorov relationship, $c = 2\sqrt{\alpha D}$, or by applying our heuristically derived fitting for Model 1, $c \approx 0.71 D^{0.64}$. Clearly the latter is more accurate for our model.

**Behaviour of $D$ as $\gamma$ becomes large**

The estimates of $D$ using Eq. (A4.6) are less accurate for larger $\gamma$. It is therefore interesting to consider the behaviour of the discrete system in the limit $\gamma \to \infty$. This system is a perturbation around the nearest neighbour kernel.

The kernel is

$$K_x = C e^{-|x|^\gamma}, \tag{A4.7}$$

The decay is very fast, such that the kernel can be reasonably represented only by the first two terms: $C \sim 1 + e^{-2\gamma}$. The diffusion constant can then be calculated:

$$D \approx \frac{1}{2C}\left(1 + 4e^{-2\gamma}\right) \approx \frac{\left(1 + 4e^{-2\gamma}\right)}{2\left(1 + e^{-2\gamma}\right)} \approx \frac{1}{2}\left(1 + 3e^{-(2\gamma - 1)}\right). \tag{A4.8}$$

This is a reasonably accurate approximation when $\gamma \geq 1.5$. A similar correction can be made in principle to the nearest neighbour wave of advance velocity, but we do not pursue this here.

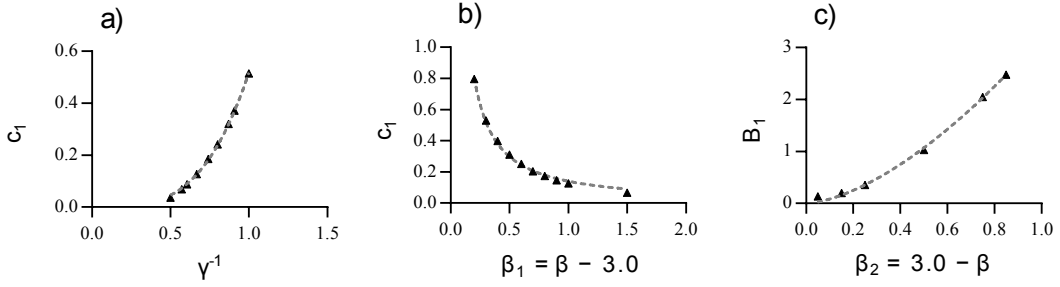

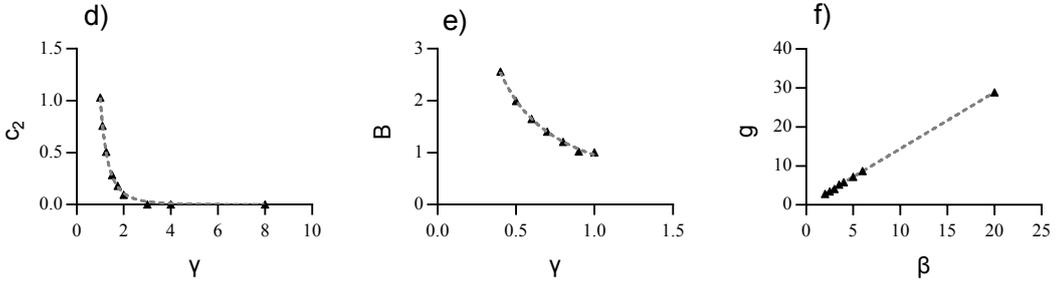

Figure A4: Deriving wave acceleration and velocity behaviour with kernel parameterisation using linear and non-linear regression. For equations, see main text. Fittings are to modified velocities $c_1 = c - 0.5$, $c_2 = c - 0.78$ and acceleration $B_1 = B = 1.0$. These fittings clarify qualitative behaviour only. a) Model 1: Exponential family kernels: $1.0 \leq \gamma \leq 2.0$; b) Model 1: Power law kernels: $3.2 \leq \beta \leq 4.5$; c) Model 1: Power law kernels: $2.15 \leq \beta \leq 2.95$; e) Model 2: Exponential family kernels: $1.0 \leq \gamma \leq 8.0$; f) Model 2: Exponential family kernels: $0.3 \leq \gamma \leq 1.0$; g) Power law kernels: $2 \leq \beta \leq 20$

## A5. Estimating maximum and minimum wave velocities for the constant velocity waves in Model 1

Model 1 corresponds well with a lattice version of the model suggested by Clark *et al* [37] with their parameter for number of offspring, $R_0 = 1$. We can therefore follow a lattice equivalent of their method for estimating minimum and maximum wave velocities based on the idea of "extreme dispersers". These are the dispersal events that travel furthest ahead of the wave front in each generation, and in doing so define both the wave front for the next generation and the velocity of the wave. The probability distribution of distances travelled by extreme dispersal events depends on occupation in the region of the lattice that can contribute the extreme disperser in a generation.

Two examples are of particular interest. Our one dimensional simulations start with a single occupied site at the far left of the lattice. In this case, occupation at the wavefront is so sparse that only one site is able to contribute the extreme disperser. The distance travelled by the wave is described simply by the dispersal kernel. Alternatively, the wavefront may be densely packed, such that occupation stretches some long distance rearward from the furthest forward occupied site. In this case, the extreme disperser could be contributed by many different sites, and the average advance of the wave in a generation will be greater. If we accept that the extreme disperser defines wave velocity and that it can only originate from a given region of the wavefront, these cases can be used to retrieve estimates of the minimum and maximum wave velocities respectively.

We begin with the case of estimating minimum velocity based on dispersal from an isolated occupied site. In the continuous case studied by Clark *et al*, the probability density function (PDF) of the extreme dispersal event with distance is

$$p(x;1) = R_0 K(x) \left[ \int_{-\infty}^{x} K(y) \mathrm{d}y \right]^{R_0 - 1} \qquad -\infty < x < \infty, \qquad (A5.1)$$

where $p(x; N_s)$ represents the probability of a single propagule travelling distance $x$ from $N_s$ evenly spaced occupied sites being the extreme disperser. $R_0$ is the number of offspring from a single occupied site per generation and $K(x)$ is the dispersal kernel. This equation corresponds to Eq. (1) of [37]. In our lattice system, we instead consider

$$p(x;1) = R_0 K_f(x) \left[ \sum_0^x K_f(y) \right]^{R_0 - 1}, \qquad (A5.2)$$

where $K_f(x)$ is the 'forward dispersal kernel' such that $K_f(0) = 0.5$, with normalisation condition $\sum_1^\infty K_f(x) = 0.5$, so as to represent the possibility of both backward and forward dispersal. The cumulative distribution function (CDF) is

$$P(x;1) = \left[ \sum_0^x K_f(y) \right]^{R_0}. \qquad (A5.3)$$

When the front is a consecutive series of occupied sites on our lattice, the CDF becomes

$$P(x; N_s) = \prod_{d=0}^{N_s} \left[ \sum_0^x K_f|y+d| \right]^{R_0}, \qquad (A5.4)$$

by which the dispersals of every site in the wavefront stretching back a distance $N_s$ travel to distance $x$ from the furthest forward occupied site or nearer. As the CDF is a discrete series, we simply obtain the PDF using

$$p(x; N_s) = P(x; N_s) - P(x - 1; N_s). \quad (A5.5)$$

The average velocity of the travelling wave is the expected dispersal distance of the extreme disperser. This is the weighted sum of the PDF,

$$E(c; N_s) = \sum_{x=0}^{\infty} p(x; N_s)x. \quad (A5.6)$$

We can use these equations to obtain minimum ($N_s = 1$) and maximum ($N_s \to \infty$) velocity estimates. Of course, practically $N_s$ cannot go to infinity in our numerical simulations, so we instead use $N_s = 10^5$, and the length of the kernel is similarly limited to $b = 10^5$. Velocity estimates were obtained through this method using Python, and were found to successfully bound the velocities retrieved from explicit simulation for those kernels that lead to constant velocity waves in Model 1. For short range kernels, the maximum velocity estimated by this method very closely matches the observed velocity. As Clark *et al* suggest, the asymptotic wave velocity of the wave of advance created by fat-tailed dispersal kernels drops away from the maximum value, though does not reach the velocity predicted by the $N_s = 1$ limit. Note that in all our models, $R_0 = 1$. An increase in $R_0$ does not correspond simply with an increase in $bN$ in Model 1 as $R_0$ does not take into account the reduced impact of each of the dispersal events. Large $R_0$ system would correspond to large $bN$ systems if carrying capacity of our sites was constrained to 1 and each full site made $N$ dispersal attempts. See Fig. 5 in the main paper for results.

### A6. Finite size scaling of a stochastic system with a bivariate Student's $t$ dispersion kernel

A kernel given by the bivariate Student's $t$ distribution $\left(\text{see Eq. (19)}, K(l) = \dfrac{1}{2\sqrt{2u}\left(1 + \dfrac{l^2}{2u}\right)^{\frac{3}{2}}}\right)$ has been discussed in ref [37]. This kernel is both convex at its source and lacks a finite second moment. It is extremely similar at long ranges to a power law kernel with $\beta = 3.0$. Unsurprisingly, we find that wave acceleration behaviour is akin to that observed for this kernel, which is explored in more detail in the main body of this study. Specifically, acceleration is marginal in the highly stochastic Model 1 when parameter $u = 1$, with a gradient on the log-log system behaviour plot, Fig. A5, of 1. Note that acceleration persists for long times when $u$ is large.

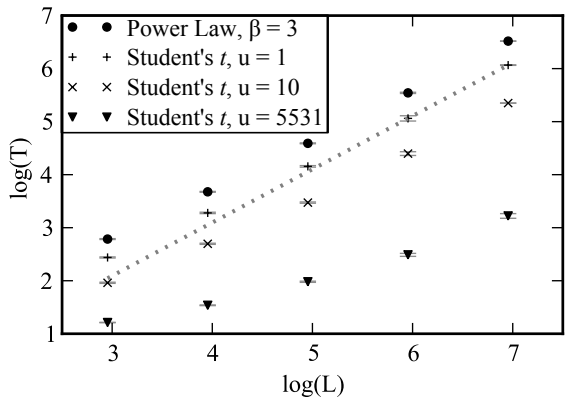

Figure A5: Model 1 comparison of finite-size scaling results with a bivariate Student's $t$ kernel and a power law kernel with $\beta = 3.0$. Note the essential similarity of the system behaviour, both tending to finite velocity over time. A linear fitting to the later stages of the simulations using a bivariate Student's $t$ kernel with parameter $u = 1$ gives a gradient of $\sim 1$, dotted line. Minimum replicates were as follows: $L = 10^3$, 50; $L = 10^4$, 50; $L = 10^5$, 20; $L = 10^6$, 5; $L = 10^7$, 1. Relative errors shown.